\documentclass[a4paper,11pt]{article}
\usepackage{jheppub}
\usepackage{dsfont}
\usepackage{amsmath,amssymb,amscd,amsfonts,mathtools}
\usepackage{graphicx}
\usepackage{hyperref}
\usepackage[dvipsnames,svgnames]{xcolor}
\usepackage[normalem]{ulem}

\preprint{UT-WI-21-2026}
\title{Holographic entanglement  entropy with conformal boundary conditions}
\author[a]{E. C\'aceres,}
\author[a]{H. Krishna,}
\author[a]{H. Palani Balaji,}
\author[a]{V. Patil}

\affiliation[a]{Theory Group, Department of Physics, University of Texas, Austin, TX 78712, USA}

\date{November 2025}
\begin{document}
 \abstract{

We study holographic entanglement entropy in 3-dimensional AdS gravity with conformal boundary conditions which fix the conformal class of the boundary metric and its extrinsic curvature, $K$, while leaving the Weyl mode dynamical. Extending Lewkowycz-Maldacena-Dong's replica construction, we derive the corresponding holographic entanglement entropy formula. We show that the fluctuating Weyl mode does not contribute additional entropy. The entropy of the full boundary is therefore the Bekenstein-Hawking entropy, and the entropy of a boundary subregion continues to obey the Ryu-Takayanagi prescription,
namely, the area of a minimal surface divided by $4G_N$.
 We carry out explicit calculations for global AdS, rotating and non-rotating BTZ geometries. For an interval in AdS$_3$   we find that the $K$-dependent holographic entanglement entropy is governed by  $c_m=\frac{3 \ell}{2 G_N}.$ For a thermal state at high conformal temperatures we find that,
  the entropy is governed by $c_{\rm eff}=\frac{3 \ell}{2 G_N} \frac{K \ell- \sqrt{K^2 \ell^2-4}}{2}$, the same effective central charge that governs the Cardy-like density of states,  in agreement with previous results in the literature. 
  
  Finally, we also  compute the entanglement entropy directly from the conjectured dual boundary theory  - a holographic CFT coupled with time-like Liouville theory and deformed by a marginal $T \bar{T}$ like operator - and find  $S_{EE} =\frac{c_{\rm eff}}{3} 
\ln\!\left(\frac{2 R \sin\phi_0}{\epsilon}\right)  .$ This result for  the state with no operator insertions (vacuum state
), provides an independent boundary realization of $c_{\mathrm{eff}}$ while clarifying that this state is not the state dual to global AdS.  }

\maketitle

\section{Introduction and summary of our results}

Quantum gravity in asymptotically anti-de Sitter (AdS) spacetime is usually formulated with Dirichlet boundary conditions. In this case, the induced metric $h_{ab}$
at the conformal boundary is held fixed $\delta h_{ab}=0$. 
Matter fields, such as scalar or gauge fields, are also assigned appropriate falloff conditions near the asymptotic boundary. With these boundary conditions, the gravitational theory admits a holographic description in terms of a conformal field theory living on the conformal boundary. This is the standard AdS/CFT correspondence \cite{Witten:1998qj,Aharony:1999ti,Maldacena:1997re} and the holographic principle \cite{Susskind:1994vu,Susskind:1998dq}. 


Dirichlet boundary conditions, however, are not always the most natural choice. In Euclidean signature, the Dirichlet problem for gravity is not elliptic and therefore does not provide a well-behaved starting point for perturbation theory \cite{Witten:2018lgb,Anderson:2006lqb}. A better-posed problem is obtained by imposing conditions on the extrinsic curvature of the boundary, for example, by requiring it to be positive or negative definite. Of particular interest are \emph{conformal boundary conditions}, in which one fixes the conformal class of the boundary metric together with the trace of the extrinsic curvature \cite{PhysRevLett.28.1082}. It has been shown that this boundary value problem is elliptic, so that perturbation theory can be defined around any solution \cite{Witten:2018lgb, Liu:2024ymn}. Related aspects of conformal boundary conditions have been studied in \cite{Anninos:2023epi,Anderson:2006lqb,Anninos:2024wpy,Anninos:2024xhc,Hamdan:2026olq,Coleman:2020jte,Galante:2025tnt} and more general boundary conditions/terms in \cite{An:2021fcq,York:1986lje,York:1986it} 
The usual AdS/CFT correspondence provides a precise map between gravitational fields, with Dirichlet boundary conditions, and CFT operators. Therefore,  it is natural to ask whether,  and how,  familiar holographic constructs change when one imposes conformal boundary conditions instead. The goal of this work is to investigate this question for holographic entanglement entropy. In this article, we will work in 3D AdS spacetime.

Recently, \cite{Allameh:2025gsa}  proposed that the holographic dual for such theories in AdS$_3$ consists of a holographic CFT coupled with a time-like Liouville theory \cite{Gibbons:1978ac,Harlow:2011ny,Schomerus:2003vv,Giribet:2011zx,Bautista:2019jau} 
and deformed by a marginal $T\bar{T}e^{-2 \xi \Phi}$ operator. The TT bar deformation has been studied in \cite{Zamolodchikov:2004ce,Smirnov:2016lqw,Cavaglia:2016oda,McGough:2016lol,Hartman:2018tkw,Donnelly:2018bef,Kraus:2018xrn}. The vertex operator $e^{-2 \xi \Phi}$ is chosen to make the $T\bar{T}e^{-2 \xi \Phi}$ operator marginal.  The authors of \cite{Allameh:2025gsa} studied the high-temperature density of states, which shows Cardy-like growth.
Thus, even though the locality of the theory remains an open question, it exhibits some thermodynamic properties characteristic of a local theory. They found that the high temperature density of states is governed by a $c_{\text{eff}}= \frac{3 \ell}{2 G_N} \frac{K \ell- \sqrt{K^2\ell^2-4}}{2}$. But there are some subtle aspects of this theory. Upon coupling the time like Liouville theory to the ordinary vanilla CFT, the total anomaly central charge $c=c_m+c_{\rm Liou}=0$ vanishes. The resulting theory is non-unitary, and there are states with $h_{\rm min}<0$. Because of the vanishing anomaly central charge, the global AdS (minimal Brown-York energy) solution is mapped to a state with an operator insertion of dimension $h_{\rm min}$.  

This motivates the question of whether the entanglement entropy of the dual theory exhibits the behavior expected of a local CFT. We will focus on entanglement entropy for subregions, since in a continuum local QFT the entropy of a spatial subregion is generically UV-divergent. Our results are as follows.
\begin{enumerate}
    \item First, we extend the bulk replica trick of Lewkowycz--Maldacena \cite{Lewkowycz:2013nqa} to gravity with conformal boundary conditions. As a warm-up application, we compute the entanglement entropy of the rotating and non-rotating BTZ black holes. The result is  the usual Bekenstein-Hawking entropy of the black hole, now obtained with conformal boundary conditions imposed at the boundary
    $$S_{BH}= \frac{A_\text{hor}}{4 G_N}.$$
    From the boundary perspective, this is the entropy of the entire system in a thermal state. Interestingly, the calculation does not require adding any counterterms to the action appropriate to conformal boundary conditions.

    \item The conformal boundary condition fixes the metric only up to a conformal class, and the trace of the extrinsic curvature. Hence, there is a fluctuating Weyl mode on the boundary, and the gravity degrees of freedom are not fully decoupled. This raises the question of whether the Ryu-Takayanagi prescription  \cite{Ryu:2006bv,Ryu:2006ef} for the entanglement entropy of a subregion at the boundary is modified (see \cite{Rangamani:2016dms} for a review). Adapting the procedures of  Dong \cite{Dong:2016fnf}  and  Casini et.al. (CHM) \cite{Casini:2011kv} we derive the RT formula for gravity with conformal boundary conditions. We find that it is still given by
    $$S_{EE}= \frac{A_\text{min-surface}}{4 G_N}.$$ Although one might have expected the fluctuating boundary Weyl mode to contribute additional terms to the entropy, we find no such contribution, and the result is again given solely by the area of the minimal surface. The dual CFT lives on the cutoff surface with a finite extrinsic curvature $K \ell>2$ having metric $ds^2= -(1+r^2/\ell^2)dt^2+r^2 d\phi^2$. For the subregion $[-\phi_0,\phi_0]$, the holographic computation gives 
    \begin{equation}
\label{bulk1}
 S_{EE}=\frac{c_m}{3} \operatorname{arcsinh}
  \left[
  \sqrt{\frac{k-\Delta}{2\Delta}}\,
  \sin\phi_0
  \right], \quad k=K \ell,\quad  \Delta= \sqrt{K^2\ell^2-4}.
\end{equation}
One can vary the extrinsic curvature and take the $K \ell \rightarrow 2^+$ limit. In this limit, the cutoff surface reaches the asymptotic boundary of AdS. Then the entropy is given by
\begin{equation}
\label{ansbulkbd}
S_{EE} =\frac{c_{m}}{3} 
\ln\!\left(\frac{  \sin\phi_0}{\epsilon_k}\right) , \quad \epsilon_k= \sqrt{\frac{\Delta}{2(k-\Delta)}}. 
\end{equation}
Here $\epsilon_k$ is the UV cutoff of the boundary theory. This is expected from the familiar UV divergence in local quantum field theory. The entanglement entropy in eq. \eqref{bulk1} has a cutoff dependence which is consistent with the effective field theory perspective. The coupling for the Liouville field $\mu \propto K \ell- 2$ seems to vanish when one takes this limit, causing it to decouple, and we are left with just matter entropy.

    \item Next, we perform a pure CFT calculation without any connection to the bulk physics. For the CFT defined on a cylinder with a radius $R$, we computed the entanglement entropy in the state with no insertion which we refer to as the vacuum state of a region $[-\phi_0,\phi_0]$. This state \textit{would not} correspond to global AdS in the bulk as emphasized earlier. The metric on the cylinder is $ds^2= -dt^2+R^2 d\phi^2$. We used the CHM map \cite{Casini:2011kv} to obtain a thermal state on the hyperbolic space with temperature as $T=\frac{1}{2 \pi R}$. The vanilla CFT is coupled to a time-like Liouville field and subsequently deformed by a marginal $T \bar{T} e^{-2 \xi \Phi}$ operator. Using the Zamolodchikov \cite{Smirnov:2016lqw} trick, we find all order deformation operators and find the Liouville saddle. Then we computed the on-shell action for the replicated manifold. Its derivative with respect to the replica index yields the entanglement entropy. 
    \begin{equation}
    \label{ansintro}
S_{EE} =\frac{c_{\rm eff}}{3} 
\ln\!\left(\frac{2 R \sin\phi_0}{\epsilon}\right)  
\end{equation}
One shouldn't compare this result with \eqref{ansbulkbd}. The state that describes the global AdS has $h_{\rm }<0$, and the entropy computation would involve the 2n-point correlator, which is also deformed by the $T \bar{T}$ deformation. In the future, we would like to return to this specific point. The motivation here was to work with the simplest possible state in the dual CFT, and glean general features of the dual theory from its entanglement entropy. 


\end{enumerate}

The content of this article is organized as follows. In section \ref{CBC-review} we review conformal boundary conditions in gravity with a negative cosmological constant. In section \ref{cone}, we use the Lewkowycz--Maldacena replica trick to compute the entanglement entropy for the non-rotating BTZ black hole and then generalize it to the rotating case. Next, in section \ref{RTderivation}, we derive the RT formula for a subregion and use it to compute the corresponding boundary subregion entropy. We then compute the entanglement entropy on the dual boundary theory for the vacuum state in section \ref{sec:chm_boundary}. Finally, in section \ref{discussion}, we conclude with a discussion and possible future directions. Technical details of the calculations are collected in Appendices \ref{calculation} and \ref{rotating_cal}. 

\section{Conformal boundary condition: A review}
\label{CBC-review}
When we consider conformal boundary conditions (CBC) \cite{Banihashemi:2024yye,Banihashemi:2025qqi,Allameh:2025gsa} 
the bulk action stays the same, but the standard boundary GHY term \cite{Gibbons:1976ue} is modified,
\begin{equation}
\label{cbc_action}
    I = -\frac{1}{16 \pi G_N}\int_{\mathcal{M}} d^3x\sqrt{g}(R -2 \Lambda)- \frac{\alpha}{8 \pi G_N}\int_{\partial\mathcal{M}} d^2x \sqrt{h} K. 
\end{equation}
Then, denoting $\bar{h}^{\mu\nu}=h^{-1/2} h^{\mu\nu}$, the variation of the action is  
\begin{eqnarray}
    \delta I \sim \int_{\mathcal{M}} EOM \times \delta g_{\mu\nu}+ \int_{\partial \mathcal{M}} \sqrt{h} \Bigg((\cdots) \delta \bar{h}_{\mu\nu}- 2 (1- \alpha) \delta K +h^{-1/2} K (1- 2\alpha ) h_{\mu\nu} \delta h^{\mu\nu}\bigg).\nonumber\\
\end{eqnarray}
For a well-defined variational principle, the conformal boundary condition entails setting

\begin{equation}\label{eq:alpha_CBC}
    \delta(h^{-1/2}h_{\mu\nu})=0, \quad \delta K = 0, \quad \alpha=1/2.
\end{equation}
For AdS$_{d+1}$, the parameter is $\alpha=1/d$. The idea here is to fix the conformal metric $\bar{h}_{\mu\nu}$ and $K$. The determinant of the $h_{\mu\nu}$  is still unfixed. The ``conjugate'' of the determinant is fixed as $\delta K=0$. Then, with this boundary condition, one can calculate the on-shell action of the gravity theory \cite{Allameh:2025gsa,Banihashemi:2025qqi}. \\

\noindent  We now review the results of \cite{Allameh:2025gsa} on the sphere partition function as a warmup. 

\subsection*{Sphere partition function}
Let us fix the conformal class of the boundary metric to be that of the round $S^2$.\footnote{The 2-sphere partition function for timelike Liouville theory has been calculated in \cite{Anninos:2021ene, Muhlmann:2022duj}.} And then we `fill' in with the bulk geometry.  
The bulk geometry that dominates the path integral is given by
\begin{eqnarray}
    ds^2= \frac{dr^2}{1+r^2}+r^2 d\Omega^2.
\end{eqnarray}
Here, we fix the extrinsic curvature of the hypersurface. We can find the extrinsic curvature at a finite radial position $r_c$, and then we can invert it to substitute $r_c$ in terms of $K$. 
\begin{eqnarray}
    K= 2 \sqrt{1+1/r_c^2}, \quad r_c= \frac{2}{\sqrt{K^2-4}}.
\end{eqnarray}
Here we choose the simple radial embedding of the boundary, for which  $n_{\mu}dx^{\mu}= dr$.\footnote{ More generally, one may allow the boundary surface to undulate along the thermal or spatial cycle; such embeddings lead to non-radial normals and additional semiclassical saddles of the gravitational partition function \cite{Hamdan:2026olq}.}
The on-shell partition function can be evaluated as
\begin{eqnarray}
    \log Z= \frac{1}{16 \pi G} \int d^3 x \sqrt{g} (R+2)+\frac{1}{16 \pi G} \int d^2 x \sqrt{h} K.
\end{eqnarray}
The bulk solution is locally $AdS_3$ and has $R=-6$, then the on-shell action becomes
\begin{eqnarray}
      \log Z= -\frac{1}{4 \pi G} \int d^3 x \sqrt{g} +\frac{K}{16 \pi G} \int d^2 x \sqrt{h} =\frac{1}{2 G} (\sinh^{-1} r_c-r_c^2\sqrt{1+1/r_c^2}+K r_c^2).
\end{eqnarray}
Using  $r_c$ in terms of $K$, we obtain,
\begin{eqnarray}
    \log Z[S^2]=\frac{\ell}{4 G} \log \frac{K \ell+2}{K \ell-2}= \frac{c_m}{6}  \log \frac{K \ell+2}{K \ell-2}
\end{eqnarray}
where we have used $c_m=3\ell/2G$ as the standard Brown-Henneaux central charge \cite{Brown:1986nw}. One can compute the Brown-York stress tensor from the action \cite{Brown:1992br}, and we get
\begin{eqnarray}
    T_{\mu\nu}= K_{\mu\nu}-\frac{1}{2} K h_{\mu\nu} \quad T_{\mu\nu}^{S^2}=0.
\end{eqnarray}

\subsection*{Torus partition function:}
Next, we study the torus partition function. This is relevant to understanding the partition function at finite temperature and in the presence of an angular potential. It can be written as the trace of the Hilbert space as
\begin{equation}
    Z(\tilde{\beta},\tilde{\Omega})= \mathrm{Tr\,\,exp}[-\tilde{\beta}(\tilde{H}-\tilde{\Omega} J)].
\end{equation}
The tilde quantity will be called a conformal quantity.
Here $\tilde{H}\equiv \lambda H^{CBC}$ is the generator of time translation of $\tilde{\tau}$ and $H^{CBC}$ is the Hamiltonian with conformal boundary condition. The potentials $\tilde{\beta},\tilde{\Omega}$ are defined as the periodicity of $\tilde{\tau}$ and $\phi$ coordinate as
\begin{eqnarray}
\label{CBC}
     &&  ds^2= \tilde{\lambda}^2(d\tau^2+R^2 d\phi^2)= \lambda^2 (d\tilde{\tau}^2+d\phi^2),\nonumber\\
     &&  (\tau,\phi)\sim (\tau+\beta,\phi+i \beta \Omega),\quad (\tilde{\tau},\phi)\sim (\tilde{\tau}+\tilde{\beta},\phi+i \tilde{\beta} \tilde{\Omega}), \quad \tilde{\beta}=\beta/R, \quad \tilde{\Omega}= R \Omega\nonumber.\\
\end{eqnarray}
With these boundary conditions, there are two contributions to the partition function: one from rotating BTZ and the other from thermal AdS. For the rotating BTZ case, the metric and corresponding periodicity of coordinates can be written as
 \begin{eqnarray}
    && ds^2= - f(r) dt^2+\frac{dr^2}{f(r)}+r^2\Big(d\theta- \frac{r_-r_+}{\ell r^2}dt\Big)^2, \quad f(r)= \frac{(r^2-r_+^2)(r^2-r_-^2)}{\ell^2 r^2}\nonumber\\
    && (t,\theta)\sim (t+i \beta, \theta+i \beta \Omega)\sim (t,\theta+2 \pi).
     \end{eqnarray}
The thermal AdS solution can be found by taking $r_-=0$ and $r_+=i \ell $ in the above solution. The inverse temperature and angular velocity can be found as
\begin{eqnarray}
    \beta= \pm \frac{4 \pi}{f'(r_{\pm})},\quad  \quad \Omega= \frac{r_{\mp}}{\ell r_{\pm}}
\end{eqnarray}
where we used $\Omega= \frac{g_{t\theta}}{g_{\theta\theta}}|_{r_{\pm}}$. At a fixed cutoff surface $r_c$, we can define a new angular coordinate as $\phi= \theta- \frac{r_-r_+}{\ell r_c^2}t$ to account for rotation at that surface. The metric on the cutoff surface is
\begin{eqnarray}
    ds^2=-f(r_c)dt^2+r_c^2 d\phi^2, \quad (t,\phi)\sim \Big(t+i \beta, \phi+i \beta (\Omega- \frac{r_-r_+}{\ell r_c^2})\Big).
\end{eqnarray}
The above metric can be written in terms of \eqref{CBC} by identifying $\tilde{\lambda}=\sqrt{f(r_c)}, R= r_c/\sqrt{f(r_c)}$ and
\begin{eqnarray}
    \tilde{\beta}=\beta/R=\beta \frac{\sqrt{f(r_c)}}{r_c}, \quad \tilde{\Omega}=\frac{r_c}{\sqrt{f(r_c)}}(\Omega- \frac{r_-r_+}{\ell r_c^2}).
\end{eqnarray}
Thus, the inverse temperature and angular velocity can be written as
\begin{eqnarray}
    \tilde{\beta}= \pm \frac{4 \pi}{f'(r_{\pm})}\frac{\sqrt{f(r_c)}}{r_c}, \quad \tilde{\Omega}=\pm \frac{r_{\mp} r_c}{r_{\pm} \ell \sqrt{f(r_c)}} (1-r_{\pm}^2/r_c^2).
\end{eqnarray}
The extrinsic curvature on the fixed hypersurface can be calculated as
\begin{eqnarray}
    K= \pm \frac{1}{r_c \sqrt{f(r_c)}}[f(r_c)+\frac{r_c f'(r_c)}{2}].
\end{eqnarray}
Essentially, we keep $K$ fixed in the path integral; hence, we often express $r_c$ in terms of $K$. 
These boundary condition implies $\tilde{\Omega}^2 \leq 1$. The black hole case is when $r_-<r_+<r_c$ and the cosmic horizon as $r_c<r_-<r_+$. The asymptotic boundary corresponds to $K \ell=2$. Then, at a fixed cutoff surface $r_c$, we have
\begin{eqnarray}
    \tilde{\beta}>0, \quad K \ell \geq 2, \quad 0 \leq \tilde{\Omega}\leq 1.
\end{eqnarray}
Using these boundary quantities, the partition function for the BTZ can be computed as
\begin{eqnarray}
 \mathrm{log\,} Z(\tilde{\beta},\tilde{\Omega})= \frac{\pi^2 \ell (K \ell- \sqrt{K^2 \ell^2-4})}{4 G \tilde{\beta}(1-\tilde{\Omega}^2)}= \frac{\mathrm{log} Z(\tilde{\beta})}{1- \tilde{\Omega}^2}  . 
\end{eqnarray}
The interesting aspect of the above equation is the appearance of the factor of $ (K \ell- \sqrt{K^2 \ell^2-4})$. Furthermore, the entropy can be calculated as
\begin{eqnarray}
    S= (1-\tilde{\beta} \partial_{\tilde{\beta}})\, \mathrm{log} Z(\tilde{\beta},\tilde{\Omega})=\frac{\pi^2 \ell (K \ell- \sqrt{K^2 \ell^2-4})}{2 G \tilde{\beta}(1-\tilde{\Omega}^2)}  =\frac{2 \pi r_{\pm}}{4 G}.
\end{eqnarray}
Having reviewed the main results of CBC in AdS$_3$, we now proceed to study the entanglement entropy in gravitational theories with these boundary conditions.  

\section{Lewkowycz-Maldacena with conformal boundary conditions}\label{cone}
In this section, we adapt the Lewkowycz-Maldacena prescription \cite{Lewkowycz:2013nqa} (see also \cite{Fursaev:1995ef}) to compute the entanglement entropy in gravitational solutions with conformal boundary conditions. This is interpreted as computing the Von Neumann entropy \eqref{ent_def} of the density matrix for the full boundary region at some time slice $t_0$, in a candidate field theory dual to the bulk with conformal boundary conditions. 

The entanglement entropy 
can be 
obtained as the limit $n\rightarrow1$ of the Rényi entropies. To calculate the Rényi entropies, we construct  $n$-times replicated manifolds with the same boundary conditions (conformal) as the original manifold. The Euclidean gravitational action can be written in terms of the partition function as
\begin{equation}
    \log Z(\beta)=-S_{\rm E,grav}.
\end{equation}
Using the partition function, we can find the thermodynamic entropy, which is then identified with entanglement entropy:
\begin{equation}
    S= -(\beta \partial_{\beta}-1) 
    \log Z.
\end{equation}
In terms of Rényi entropy, we have,
\begin{equation}
\label{ent_def}
    S = -n \partial_n \left[ \log Z(n)-n\log Z(1) \right]_{n=1}= -Tr[\hat{\rho} \log \hat{\rho}],
\end{equation}
where $\log Z(n)=Tr[\rho^n]$ is the action of the n-replicated manifold and $\hat{\rho}=\frac{\rho}{Tr[\rho]}$ is the properly normalized density matrix. The replicated geometry has the period of the $\tau$ circle to be $\tau \sim \tau+2\pi n$. In particular, we consider Euclidean solutions with a $U(1)$ symmetry. In these time translation invariant solutions, the action of the replicated manifold is n times the action integrated over $\tau$ from 0 to $2\pi$:
\begin{equation}
    \log Z[n]= n [\log Z[n]]_{2 \pi}.
\end{equation}
Here $[\log Z[n]]_{2 \pi}$ means the gravitational action of solution labeled by $n$ but the $\tau$ is integrated from $[0, 2 \pi]$. With these simplifications, the entropy becomes
\begin{equation}
    S = -n^2\partial_n \left[ \log Z(n) \right]_{2\pi}.
\end{equation}
We can think of this term as the gravitational action of the manifold with periodicity $\tau \sim \tau+2\pi$ but with a conical singularity at $r=0$ with an opening angle $2\pi/n$. The derivative with respect to (wrt) $n$ can be written in terms of the variation of the metric wrt the fields \footnote{ The derivative with $n$ means we need to change $n$ slightly. And changing n changes the opening angle and the strength of the singularity. Hence, the metric and other fields will change. The
solution with $n=1$ is a solution of the equations of motion. Then the first order variation around $n=1$ will vanish due to the equation of motion. Thus, the change in the action comes due to the boundary term.}
\begin{equation}
\begin{split}
    &-\partial_n \left[ \log Z(n) \right]_{2\pi} |_{n=1}  = -\sum_i \frac{\delta S_n}{\delta \phi_i} \partial_n\phi_i\\
    & = \int_{\mathcal{M}} (\text{EOM}) \partial_ng + \frac{1}{16\pi G_N} \int_{\partial\mathcal{M}} \sqrt{h} n^\mu g^{\nu\gamma} \left( \nabla_\gamma (\partial_n g_{\mu\nu}) - \nabla_\mu (\partial_n g_{\nu\gamma}) \right)\\
    & = 0 + \frac{2 \pi}{16 \pi G_N} \int_{r=0} 2 dx,
\end{split}    
\end{equation}
where in the second line, the boundary term arises from the variation of the action. At the large r boundary this term cancels with the variation of the GHY term present in the action. But the conical singularity at $r=0$ acts as another boundary and no such term is present in the action to cancel it. This gives the only non-zero term of the entropy in terms of the transverse area (equivalently the area of the horizon):
\begin{equation}
    S = \frac{A_{H}}{4 G_N}.
\end{equation}

This is the standard derivation of entanglement entropy of a spacetime with the Einstein-Hilbert action (along with the GHY boundary term) and Dirichlet boundary conditions: $\delta h_{\mu\nu} = 0$.

When we consider\textbf{ conformal boundary} conditions, the bulk action remains the same. However,  as shown in \eqref{eq:alpha_CBC}, an additional factor of $1/d$ is present in the boundary term. In $AdS_3,$ $d=2$, we have, 

\begin{equation}
    I = -\frac{1}{16 \pi G_N}\int_{\mathcal{M}} d^3x\sqrt{g}(R-2 \Lambda) - \frac{1}{16 \pi G_N}\int_{\partial\mathcal{M}} d^2x \sqrt{h} K, 
\end{equation}
and the boundary conditions can be written as:
\begin{equation}
    \delta(h^{-1/d}h_{\mu\nu})=0, \quad \delta K = 0,
\end{equation}
thus fixing the conformal metric and  the trace of the extrinsic curvature at the boundary.

Calculating the entropy of  $\mathcal{M}$ in this case gives us the same final result in terms of area of horizon. This is to be expected since the ``boundary" at $r=0$ only arises from the conical singularity of the replicated manifold and is not the physical boundary of the spacetime. So the change in the GHY term when using conformal boundary conditions only affects the large-$r$ boundary and the term at $r=0$ which arises deep in the bulk stays the same. This calculation is explicitly done as follows:
\begin{equation}
\begin{split}
    \delta I &= -\frac{1}{16\pi G_N}\int_{\mathcal{M}}\delta[\sqrt{g}(R- 2 \Lambda)] -\frac{1}{16 \pi G_N}\int_{\partial\mathcal{M}}\delta(\sqrt{h}K)\\
    & = \int_\mathcal{M}(\text{EOM})\delta g_{\mu\nu} +\frac{1}{16\pi G_N} \int_{r=0} \sqrt{h} (2\delta K +K^{ij}\delta g_{ij}),\\
\end{split}
\end{equation}
where $i,j$ are the coordinates of the boundary manifold. Hence, the on-shell entropy becomes:
\begin{equation}
\label{area_cbc}
\begin{split}
    S &= -\partial_n \left[ \log Z(n) \right]_{2\pi} |_{n=1}  \\
    & = 0 + \frac{1}{16\pi G_N} \int_{r=0} \sqrt{h} (2\partial_n K + K^{ij}\partial_n g_{ij}) \\
    &=\frac{2 \pi}{16 \pi G_N} \int_{r=0} 2 dx \\
    & = \frac{A_H}{4G_N}.
\end{split}
\end{equation}
In the second last line, the factor of $2 \pi$ appears due to the $\tau$ integration (see more detailed exposition in section \ref{RTderivation}). 

\subsection{The BTZ black hole }
We now compute, using the Lewkowycz-Maldacena prescription, the entanglement entropy for the BTZ black hole background \cite{Banados:1992wn,Banados:1992gq}, for both the non-rotating and rotating cases. The key ingredient in achieving this is the \emph{classical approximation}, which states that in the bulk prescription, $-\log\,Z(1)$ and $-\log\,Z(n)$ have the interpretation of computing the Euclidean Einstein-Hilbert action for the black hole geometry with boundary periodicity $\beta$ and $n\,\beta$ respectively.  

It is important to note that for the classical approximation to be meaningful, the replicated manifold for $n>1$ constructed this way should still be a saddle point solution of Einstein's equation. This means that the bulk manifold should be free from conical singularities since nothing in the physical stress tensor can source these singularities.


\subsection*{Non-rotating BTZ}

The  metric for a non-rotating, Euclidean BTZ black hole is given by:
\begin{equation}
    ds^2=f(r) d\tau^2+\frac{1}{f(r)}dr^2 +r^2d\phi^2,\quad f(r)=r^2-r_H^2,
\end{equation}
where the AdS curvature scale, $\ell$ has been set to one. The topology of the boundary is that of a torus consisting of the $\tau$-circle and the $\phi$-circle.


The Hawking temperature is given by the periodicity of the $\tau$-circle at the horizon. In the near-horizon region ($r\rightarrow r_H+\epsilon$), $f(r)\approx2 r_H\epsilon$, which renders the $r$-$\tau$ part of the metric as follows:
\begin{equation}
\label{nonrot-nhr}
    ds^2\approx2 r_H\epsilon\,d\tau^2+\frac{1}{2 r_H\epsilon}d\epsilon^2.
\end{equation}
Now solving for a new coordinate $\sigma$, such that:
\begin{equation}
    \frac{1}{2 r_H\epsilon}d\epsilon^2=d\sigma^2,
\end{equation}
gives $\sigma=\sqrt{2\epsilon/r_H}$, and \eqref{nonrot-nhr} becomes:
\begin{equation}
\label{nonrot-nhr-2}
    ds^2\approx d\sigma^2+\sigma^2r_H^2 d\tau^2.
\end{equation}
Demanding that the metric be regular at the horizon fixes the periodicity of the $\tau$-circle to be $\beta=2\pi/r_H.$ 



Now we need to obtain the replicated geometry. The idea is to promote the periodicity of the $\tau$-circle to $n$ times its original value while retaining a smooth geometry at the horizon. From \eqref{nonrot-nhr-2}, it is clear that the former can be achieved by making the substitution $r_H\rightarrow r_Hn^{-1}$. It is easy to see that this gives a smooth geometry in the vicinity of the horizon by going to a new radial coordinate $\rho^2=f(r)$, which renders the metric as follows:
\begin{equation}\label{lm}
    ds^2=\rho^2d\tau^2+\frac{d\rho^2}{\rho^2+r_H^2}+(\rho^2+r_H^2) d\phi^2.    
\end{equation}
For the replicated manifold, the metric in the $\tau-\rho$ directions near the horizon is
\begin{equation}
    ds^2\approx \rho^2 d\tau^2 +n^2\frac{d\rho^2}{r_H^2}.
\end{equation}
This can be written as $ds^2=n^2(\rho^2d\tilde{\tau}^2+d\rho^2)/r_H^2$ with $\tilde{\tau}=r_H\tau/n$, which is smooth as $\rho\rightarrow0$.\\
Hence, the $n$-replicated manifold is then described by the metric:
\begin{equation}
   \label{btzreplicated}
  ds^2=f(r)d\tau^2+\frac{dr^2}{f(r)}+r^2d\phi^2,\quad f(r)=\left(r^2-\frac{r_H^2}{n^2}\right).
\end{equation}
This is by no means a unique way to implement the replica trick. See \cite{Lewkowycz:2013nqa} for a discussion of other equivalent prescriptions. 

We now proceed to calculate the entanglement entropy.
Let the boundary be characterized by a  
radial coordinate $r_c>r_H$. The induced metric and the trace of the extrinsic curvature at this hypersurface is given by:

\begin{eqnarray}
   ds^2|_{r=r_c}&&=\left(r_c^2-\frac{r_H^2}{n^2}\right)d\tau^2+r_c^2d\phi^2,\\
   K&&= \frac{2n^2 r_c^2-r_H^2}{n^2 r_c\sqrt{\frac{n^2r_c^2-r_H^2}{n^2}}}.
\end{eqnarray}
The on-shell action \eqref{cbc_action} for this replicated geometry evaluates to:
\begin{equation}
   I_n= -\frac{1}{4 G}\frac{\pi r_H}{n},
\end{equation}
where factors of $n$ enter the action explicitly through the periodicity of $\tau$ and the limits of integration of $r$, which now goes from $r_H/n$ to $r_c$.

Then the entanglement entropy of the boundary region as follows:
\begin{equation}
\label{cbc_ent}
    S = n\partial_n \left[I_n-n\,I_1 \right]_{n=1}= \frac{2 \pi r_h}{4 G_N},
\end{equation}
which is the usual Bekenstein-Hawking formula for the black hole\footnote{Note that $I_1$ is the action evaluated on the n=1 manifold which is also just $I_{n\rightarrow1}$. For technical details on analytically continuing n beyond integers, see \cite{Lewkowycz:2013nqa}.}. In the action \eqref{cbc_action}, there are no counterterms. Using that, we found the finite action and the EE for the BTZ case.  We emphasize here that this is an \textbf{exact} expression for the entropy, unlike the leading term in a large boundary expansion for the case of gravity with Dirichlet boundary conditions. Moreover, since \eqref{cbc_ent} is independent of $r_c$, the entropy for the full boundary region (thermal case) is invariant up to the Weyl class of boundary metrics, or equivalently up to the location of the conformal boundary. 

For later use, we express the entropy in terms of the boundary data as follows: 
\begin{equation}
\boxed{
    S=\frac{\pi^2 \ell}{2 G_N \tilde{\beta}}( K \ell -\sqrt{K^2 \ell^2- 4})}
\end{equation}
where the inverse conformal temperature $\tilde{\beta}$ is defined below, and $\ell$ is re-introduced in the final result. 

\subsection*{A note on conformal temperature}

There are three notions of temperature relevant here. The inverse Hawking temperature, denoted by $\beta$, is fixed by regularity of the Euclidean black hole geometry and is defined with respect to the time coordinate normalized at the asymptotic AdS boundary.  At a finite radius hypersurface, one can define a proper redshifted inverse temperature, $\beta_p= \beta \sqrt{g_{\tau \tau}}$ or a proper conformal (inverse) temperature as introduced in \cite{Allameh:2025gsa}:

\begin{equation}
    \Tilde{\beta}= \beta \frac{\sqrt{g_{\tau\tau}}}{\sqrt{g_{\phi\phi}}},
\end{equation}
which accounts for the redshift along both time and space components.

For the n-replicated manifold in coordinates $\rho$-$\tau$-$\phi$ \eqref{lm},
this relation becomes:
\begin{equation}
    \Tilde{\beta}=\frac{\rho_c}{\ell\sqrt{r_H^2 n^{-2}+\rho_c^2}}\beta.
\end{equation}
This redefinition of the inverse temperature can be identified from a conformal transformation at the boundary:
\begin{eqnarray}
    ds^2|_{\rho=\rho_c}&& =\rho_c^2d\tau^2+(\rho_c^2+r_H^2 n^{-2}) d\phi^2\longrightarrow (\rho_c^2+r_H^2 n^{-2})(d\Tilde{\tau}^2+d\phi^2),\\
    \Tilde{\tau}&&=\frac{\rho_c}{\sqrt{r_H^2n^{-2}+\rho_c^2}}\tau,
\end{eqnarray}
which evidently only affects the boundary term in the evaluation of the on-shell action.
\subsection{Rotating BTZ}
We now repeat the procedure above to calculate the entanglement entropy for a rotating BTZ black hole background. 
The Euclidean metric for $r\in[r_+,r_c]$ is given by \cite{Allameh:2025gsa}\footnote{Despite the factor of $i$ in the $d\tau d\theta$ cross-term, the metric has three positive eigenvalues.}:
\begin{equation}
\label{rotbtz}
    ds^2=f(r)d\tau^2+\frac{dr^2}{f(r)}+r^2(d\theta-i\frac{r_-r_+}{r^2}d\tau)^2,\quad f(r)=\frac{(r^2-r_+^2)(r^2-r_-^2)}{r^2},
    \end{equation}
where once again the AdS scale,  $\ell$ is set to one. At the conformal boundary $r=r_c$, by defining a new angular coordinate $\phi=\theta-\frac{r_-r_+}{r_c^2}t$, it is easy to see that the boundary topology is that of a twisted torus. While the angular velocity of the black hole is given by $\Omega=r_-/( r_+)$, switching to the coordinate $\phi$ has the effect of shifting the angular velocity, $\Omega\rightarrow\Omega-r_-r_+/(r_c^2)$, which has the effect of subtracting the rotation at $r_c$.

The Hawking temperature is obtained by looking at the metric in the near outer-horizon region:
\begin{equation}
\label{nhr-rot}
    f(r)=\frac{2(r_+^2-r_-^2)(r-r_+)}{r_+}+\mathcal{O}((r-r_+)^2).
\end{equation}
Proceeding as before, we look for a coordinate $\sigma$ such that:
\begin{equation}
    d\sigma^2=\frac{1}{f(r)}dr^2\implies\sigma=\sqrt{\frac{2r_+(r-r_+)}{(r_+^2-r_-^2)}},
\end{equation}
which renders the $\tau$-$r$ part of the near-horizon metric as follows:
\begin{equation}
    ds^2=d\sigma^2+\sigma^2\frac{(r_+^2-r_-^2)^2}{r_+^2}d\tau^2.
\end{equation}
The periodicity of the $\tau$-circle is fixed by demanding that there be no conical singularity at the horizon:
\begin{equation}
\label{beta}
    \int_0^{\beta}\frac{(r_+^2-r_-^2)}{r_+}\,d\tau=2\pi\implies\beta=2\pi \frac{r_+}{r_+^2-r_-^2}.
\end{equation}

To achieve the correct $\tau$-periodicity for the n-replicated manifold, we now scale both the inner and outer horizons as follows:
\begin{equation}
    r_{-}\rightarrow\frac{r_-}{n},\quad r_+\rightarrow \frac{r_+}{n}.
\end{equation}
It is straightforward to check that \eqref{rotbtz} with the above substitution is still a solution to Einstein's equation with $R=-6$. Here, we explicitly show that no conical singularity is introduced at the horizon by going to the coordinates $\rho$-$\tau$-$\phi$ in which the  metric near the outer-horizon region becomes:
\begin{equation}
\label{rho_rot}
    ds^2\approx\rho^2d\tau^2+n^2\,\frac{r_+^2}{(r_+^2-r_-^2)^2}d\rho^2.
\end{equation}
By redefining $\tau$ to $\tilde{\tau}=(r_+^2-r_-^2)\tau/(n r_+)$ with a periodicity of $2\pi$, the near-horizon metric becomes $ds^2=n^2r_+^2(\rho^2d\tilde{\tau}^2+d\rho^2)/(r_+^2-r_-^2)^2$ which is smooth as $\rho\rightarrow0$.

The metric for the replicated manifold is,
\begin{equation}
   \label{rotbtzreplicated}
  ds^2=f(r)d\tau^2+\frac{dr^2}{f(r)}+r^2(d\theta-i\frac{r_-r_+}{n^2r^2}d\tau)^2,\quad f(r)=\frac{(r^2-\frac{r_+^2}{n^2})(r^2-\frac{r_-^2}{n^2})}{r^2}.\\
\end{equation}
The induced metric and the trace of the extrinsic curvature of a constant-$r_c$ hypersurface in the geometry \eqref{rotbtzreplicated}, for $r_+<r_c$ are given by,
\begin{eqnarray}
   ds^2|_{r=r_c}&&=(r_c^2-\frac{r_-^2+r_+^2}{n^2})d\tau^2-2i\frac{r_-r_+}{n^2}d\tau d\theta+r_c^2d\theta^2,\\
   K&&= \frac{2n^2 r_c^2-r_-^2-r_+^2}{n^2 r_c\sqrt{\frac{(n^2\,r_c^2-r_-^2)(n^2\,r_c^2-r_+^2)}{n^4 r_c^2}}},\nonumber
\end{eqnarray}
which immediately renders the on-shell action, and subsequently the Von Neumann entropy of the boundary in the classical approximation as follows:
\begin{equation}
   I_n= -\frac{1}{4 G}\frac{\pi r_+}{n}\implies S=\frac{2 \pi r_+}{4 G_N}.
\end{equation}
In boundary variables, the result becomes:
\begin{equation}
\boxed{
    S_{\text{ent}}=\frac{\pi^2 \ell}{2 G \tilde{\beta}}\frac{( K \ell -\sqrt{K^2 \ell^2- 4})}{1- \tilde{\Omega^2}}},
\end{equation}
where the boundary potentials are given by:
\begin{equation}
    \tilde{\beta}=\beta\frac{\sqrt{f(r_c)}}{r_c},\quad\tilde{\Omega}=\frac{r_c}{\sqrt{f(r_c)}}\left(\Omega-\frac{r_-r_+}{\ell r_c^2}\right).
\end{equation}

So far, we have established that the entanglement entropy for the full boundary in a bulk black hole solution with conformal boundary conditions is still governed by the familiar Bekenstein-Hawking formula. It is then natural to ask how the Ryu-Takayanagi prescription for entanglement entropy of \emph{subregions} is modified in the presence of conformal boundary conditions. Since minimal surfaces in the bulk will now be anchored to the conformal boundary, one can expect the Weyl mode at the boundary to enter the expressions of entropy through the parameter $r_c$. This is the subject of discussion for the next section.



\section{Conformal boundary conditions and subregion entanglement entropy}
\label{RTderivation}
The RT formula has reshaped our understanding of entanglement in quantum systems \cite{Ryu:2006bv,Ryu:2006ef,Hubeny:2007xt}. It states that the entanglement entropy of a boundary subregion is given by the area of the minimal bulk surface homologous to that subregion and anchored on its entangling surface. The RT prescription was originally formulated for subregions defined on the asymptotic boundary of AdS. In our setup, however, conformal boundary conditions place the boundary at a finite radial position. It is therefore essential to derive the corresponding RT formula in this finite-boundary setting. Some work has been done in this context in \cite{Grado-White:2020wlb,Park:2018snf}.


We begin by reviewing Dong’s argument \cite{Dong:2016fnf} (see also \cite{Fursaev:1995ef,Fursaev:2013fta}), which gives an expanded version of the Lewkowycz-Maldacena construction \cite{Lewkowycz:2013nqa}, for the entanglement entropy of subregions. This argument can be generalized to the case of conformal boundaries because the RT surface arises as the fixed-point locus of the $\mathbb{Z}_n$ orbifold. The presence of a Weyl mode does not change the argument. We then use the CHM map \cite{Casini:2011kv} to compute the entanglement entropy of subregions in this setting.

First, let us review the Dirichlet case and consider a subregion in Euclidean $AdS_3$ with  radius $\ell.$
The replica construction gives a branched cover
\(M_n\) of the boundary spacetime. Next, we need to find the corresponding bulk saddle \(B_n\) with boundary \(M_n\). In the leading order
\[
  Z[M_n]=\exp[-I_{\rm bulk}[B_n]] .
\]
We assume the \(\mathbb Z_n\) replica symmetry is unbroken, one may quotient the
smooth parent geometry by this symmetry and write
\(\widehat B_n\equiv B_n/\mathbb Z_n\).  For integer \(n\), the quotient has a
codimension-two fixed surface, and we have
\[
  I_{\rm bulk}[B_n]=nI_{\rm bulk}[\widehat B_n].
\]
This follows from the locality of the bulk action.
Dong proposed this in terms of cosmic brane arguments. The quotient \(\widehat B_n\) is treated
as a geometry with a conical defect. The other perspective is to use a codimension-two brane \(C_n\), with
tension \(T_n=(n-1)/(4nG_N)\).  Varying the on-shell Einstein action with
respect to \(n\) localizes the variation on a small tube around \(C_n\) (see \eqref{eq:dong-DBC}). 
The Rényi entropy is defined as 
\begin{eqnarray}
    S_n=\frac{1}{1-n}(\log Z[M_n]- n \log Z[M_1]).
\end{eqnarray}
Using bulk geometry, we have 
\begin{eqnarray}
    S_n=\frac{n}{n-1}(I_{\rm bulk}[\hat{B}_n]- I_{\rm bulk}[\hat{B}_1]).
\end{eqnarray}
Then, we have
\begin{eqnarray}
    \partial_n \left[\frac{n-1}{n}S_n\right]= \partial_n I_{\rm bulk}[\hat{B}_n].
\end{eqnarray}
The RHS can be evaluated explicitly. The bulk solution $\hat{B}_n$ can be thought of as a family of solutions, but with varying boundary conditions near the brane. The variation of the action gives a boundary term near the brane, and we have 
\begin{equation}
\partial_n I_{\rm bulk}[\hat{B}_n]=\int \frac{d^d x \sqrt{\gamma}}{16 \pi G_N}  \widehat n^\mu
  \left(
    \nabla^\nu\partial_nG_{\mu\nu}
    -
    G^{\nu\alpha}\nabla_\mu\partial_nG_{\nu\alpha}
  \right).
  \label{eq:dong-DBC}
\end{equation}
Here $ G_{\mu\nu}$ is the bulk metric and $x$ are the coordinates on the tube, and $\gamma$ is the corresponding induced metric on the tube. The outward unit normal to the brane is denoted as $\widehat n^\mu$. One can use polar coordinates near the brane, and the metric on the tubular neighborhood can be written as
\begin{eqnarray}
    ds^2= dr^2+\frac{r^2}{n^2} d\phi^2+g_{ij}dy^i dy^j.
\end{eqnarray}
Here, the polar coordinate $\phi$ has period $2\pi$ and $y^i$ are the coordinates on the brane. We can evaluate \eqref{eq:dong-DBC} for the tubular metric, and we have
\begin{equation}
 \partial_n \left[\frac{n-1}{n}S_n\right]=\partial_n I_{\rm bulk}[\hat{B}_n]=\int \frac{d^d x \sqrt{\gamma}}{16 \pi G_N}
  \frac{2}{nr}=\frac{{\rm Area}(C_n)}{4G_N}.
  \label{eq:dong-DBC1}
\end{equation}

This
gives the refined Rényi area law
\[
  \widetilde S_n
  \equiv
  n^2\partial_n\left[\frac{n-1}{n}S_n\right]
  =
  \frac{{\rm Area}(C_n)}{4G_N}.
\]
At \(n=1\), the brane tension vanishes and \(C_1\) becomes the usual RT
surface.\\


We now turn  to our main case of interest,   \emph{ conformal boundary conditions.}  In this case we   fix the conformal class of the metric and the trace of the extrinsic curvature.
In three bulk dimensions, the action appropriate to this variational problem is
\begin{equation}
  I_{\rm CBC}
  =
  -\frac{1}{16\pi G_N}
  \int_{\mathcal M} d^3x\sqrt G\,\left(R+\frac{2}{\ell^2}\right)
  -\frac{1}{16\pi G_N}
  \int_{\partial\mathcal M} d^2x\sqrt h\,K .
  \label{eq:cbc-action}
\end{equation}
The boundary term has one-half of the usual Dirichlet Gibbons-Hawking-York
coefficient.  We write
\[
  h_{ab}=e^{2\Phi}\widetilde h_{ab},
  \qquad
  \delta\widetilde h_{ab}=0,
  \qquad
  \delta{K}=0,
\]
Equivalently, one fixes \(h^{-1/d}h_{ab}\), with \(d=2\),
and leaves the Weyl mode \(\Phi\) dynamical.  The replica manifold must
therefore be constructed in the CBC ensemble. The conformal temperature and extrinsic curvature also depend on $n$.

\subsection{Dong argument generalized for CBC}
\label{sec:dong-argument-cbc}

Let \(A\) be a boundary interval in the fixed conformal frame
\(\widetilde h_{ab}\).  We keep the notation \(B_n\) for the dominant bulk
saddle, but now \(B_n\) is the solution of the CBC variational problem rather
than the Dirichlet one.  The replica problem is defined by holding fixed the
replicated conformal boundary metric and extrinsic curvature. Thus
\begin{equation}
  Z_{\rm CBC}[M_n;\widetilde h_n,K]
  =
  \exp\!\left[-I_{\rm CBC}[B_n]\right].
  \label{eq:dong-12-cbc}
\end{equation}
The replica solution has conformal period \(\widetilde\beta_n=n \beta\) and conformal boundary condition gives $\delta K(n)=0$. Now, we deviate from Dong's construction and explicitly construct the replica geometry, computing the Rényi and Von Neumann entropies. We will work in 3-dimensional AdS.\\

We start with a static \(U(1)_\tau \times \mathbb R_u\)-symmetric
bulk geometry. The most general static ansatz with these symmetries (in radial gauge $g_{uu}=\rho^2$) is \footnote{The geometry we considered above also needs to be time reflection symmetric. These are not the most general ansatz. But this is much closer to the AdS-Rindler geometry, which is the relevant bulk dual to a subregion of the CFT vacuum state. In any case, the geometry near the brane can always be written as \eqref{eq:dong-cbc-local-cone-new}. This gives the RT formula. }
\begin{equation}
  ds^2
  =
  A(\rho)d\tau_E^2+\frac{d\rho^2}{B(\rho)}+\rho^2du^2 .
  \label{eq:dong-cbc-AB-ansatz-new}
\end{equation}
The Einstein tensor is
\[
  E_{\mu\nu}\equiv R_{\mu\nu}+\frac{2}{\ell^2}G_{\mu\nu}.
\]

For \eqref{eq:dong-cbc-AB-ansatz-new}, two independent combinations of the
Einstein equations are
\begin{equation}
  \frac{E_{\tau\tau}}{A}-B E_{\rho\rho}
  =
  \frac{AB'-BA'}{2\rho A},
  \qquad
  -\frac{2\ell^2 A}{\rho}E_{uu}
  =
  \ell^2(AB'+BA')-4\rho A .
  \label{eq:dong-cbc-AB-einstein-new}
\end{equation}
Setting \(E_{\tau\tau}=E_{\rho\rho}=0\) gives \(A/B=\alpha^2\), with
\(\alpha\) constant.  Thus \(A=\alpha^2B\).  Substituting this into
\(E_{uu}=0\) gives
\[
  B'(\rho)=\frac{2\rho}{\ell^2},
\]
so
\begin{equation}
  B(\rho)=\frac{\rho^2}{\ell^2}-\mu .
  \label{eq:dong-cbc-B-solution-new}
\end{equation}
Writing \(\mu=\rho_h(n)^2/\ell^2\), and one can set the constant \(\alpha=1\) without loss of generality. The local AdS solution becomes
\begin{equation}
  ds_n^2
  =
  f_n(\rho)d\tau_E^2+\frac{d\rho^2}{f_n(\rho)}+\rho^2du^2,
  \qquad
  f_n(\rho)=\frac{\rho^2-\rho_h(n)^2}{\ell^2}.
  \label{eq:dong-cbc-adapted-metric-new}
\end{equation}

In the gauge  above, a
 horizon segment is described by the endpoints \(u_1,u_2\), and only the
proper length
\(\int\sqrt{\gamma_{uu}}\,du\) is invariant. If we  define $ L_u\equiv\int_{u_1}^{u_2}du=u_2-u_1,$ the physical length of this horizon segment is
\begin{equation}
A_h(n)=\int_{C_n}\sqrt{\gamma_n}=\rho_h(n)L_u .
  \label{eq:dong-cbc-horizon-area-new}
\end{equation}

Near \(\rho=\rho_h(n)\), regularity of the smooth parent saddle fixes the
ordinary period
\begin{equation}
  \beta_n=\frac{2\pi\ell^2}{\rho_h(n)} .
  \label{eq:dong-cbc-beta-new}
\end{equation}
The conformal boundary has fixed $K$, so it lies on a fixed cutoff surface.
At the cut off surface \(\rho=\rho_c(n)\), the induced metric is
\begin{equation}
  ds_{\partial}^2
  =
  f_n(\rho_c)d\tau_E^2+\rho_c(n)^2du^2
  =
  \rho_c(n)^2\left(d\widetilde\tau^{\,2}+du^2\right),
  \qquad
  \widetilde\beta_n
  =
  \beta_n\frac{\sqrt{f_n(\rho_c)}}{\rho_c(n)} .
  \label{eq:dong-cbc-conformal-period-new}
\end{equation}
The trace of the extrinsic curvature of the cutoff surface is
\begin{equation}
  K(n)
  =
  \frac{2\rho_c(n)^2-\rho_h(n)^2}
  {\ell\,\rho_c(n)\sqrt{\rho_c(n)^2-\rho_h(n)^2}} .
  \label{eq:dong-cbc-wall-K-new}
\end{equation}
Redefining the variables as
\[
  k\equiv K\ell,
  \qquad
  \Delta\equiv\sqrt{k^2-4}, \quad \widetilde\beta_n=2\pi n,
\]
Now we can solve $\rho_h(n)$ and $\rho_c(n)$ in terms of conformal temperature $\widetilde\beta_n$ and $K$. Upon solving, we have
\begin{equation}
  \rho_h(n)
  =
  \frac{\ell}{2n}(k-\Delta),
  \qquad
  \rho_c(n)^2
  =
  \frac{\ell^2}{2n^2}\frac{k-\Delta}{\Delta}.
  \label{eq:dong-cbc-rhoh-rhoc-new}
\end{equation}
It is important to note that the factor \(k-\Delta\) enters through the horizon
radius, which is similar in spirit to eq 2.20 of \cite{Allameh:2025gsa}. Now, we evaluate the action of the replicated manifold. The solution has \(R=-6/\ell^2\) and \(\sqrt g=\rho\).

\begin{align}
  I_{\rm bulk}^{(n)}
  &=
  -\frac{1}{16\pi G_N}
  \int d^3x\sqrt g\left(R+\frac{2}{\ell^2}\right)
  \nonumber\\
  &=
  \frac{\beta_nL_u}{8\pi G_N\ell^2}
  \left(\rho_c(n)^2-\rho_h(n)^2\right).
  \label{eq:dong-cbc-bulk-action-new}
\end{align}
For the boundary term, one has
\[
  \sqrt h\,K
  =
  \frac{2\rho_c(n)^2-\rho_h(n)^2}{\ell^2},
\]
and we have
\begin{equation}
  I_{\rm GHY}^{(n)}
  =
  -\frac{\beta_nL_u}{16\pi G_N\ell^2}
  \left(2\rho_c(n)^2-\rho_h(n)^2\right).
  \label{eq:dong-cbc-wall-action-new}
\end{equation}
The total sum of the action becomes
\begin{equation}
  I_n
  =
  I_{\rm bulk}^{(n)}+I_{\rm GHY}^{(n)}
  =
  -\frac{\beta_nL_u\,\rho_h(n)^2}{16\pi G_N\ell^2}
  =
  -\frac{L_u\rho_h(n)}{8G_N}.
  \label{eq:dong-cbc-total-action-new}
\end{equation}
Using \eqref{eq:dong-cbc-rhoh-rhoc-new},
\begin{equation}
  \log Z_n=-I_n
  =
  \frac{\ell L_u}{16G_N}\,
  \frac{k-\Delta_K}{n}.
  \label{eq:dong-cbc-logZ-new}
\end{equation}
The von Neumann entropy is obtained by the replica derivative
\begin{align}
  S_A
  &=
  \left(1-n\partial_n\right)\log Z_n\big|_{n=1}
  \nonumber\\
  &=
  \frac{\ell L_u}{8G_N}(k-\Delta_K)
  =
  \frac{\rho_h(1)L_u}{4G_N}
  =
  \frac{A_h(1)}{4G_N}.
  \label{eq:dong-cbc-area-law-new}
\end{align}
This is the expected horizon-area answer.


We can also do a calculation similar to Dong's tubular computation \eqref{eq:dong-DBC} as reviewed in the previous subsection. 
 Excise a tube \(T_\epsilon\) of radius \(R=\epsilon\) around the
brane.  In local coordinates near the brane (or fixed point of the orbifold)
\begin{equation}
  ds^2
  =
  dR^2+\frac{R^2}{n^2}d\tau^2+\rho_h(n)^2du^2+\cdots,
  \qquad
  \tau\sim\tau+2\pi .
  \label{eq:dong-cbc-local-cone-new}
\end{equation}
Here $u$ is a coordinate along the brane.
For the variation with respect to \(n\), the singular part comes from
\[
  \partial_nG_{\tau\tau}
  =
  \partial_n\left(\frac{R^2}{n^2}\right)
  =
  -\frac{2R^2}{n^3}.
\]
The \(n\)-dependence of \(\rho_h(n)^2du^2\) is finite at the tube and gives no
leading contribution as \(\epsilon\to0\).  The small-tube boundary integrand is (eq. 17 of Dong \cite{Dong:2016fnf})
\begin{equation}
  \widehat n^\mu
  \left(
    \nabla^\nu\partial_nG_{\mu\nu}
    -
    G^{\nu\rho}\nabla_\mu\partial_nG_{\nu\rho}
  \right)
  =
  \frac{2}{nR}+O(R^0).
  \label{eq:dong-cbc-tube-integrand-new}
\end{equation}
Since \(\sqrt\gamma=(R/n)\rho_h(n)\) on the tube, the localized contribution is
\begin{align}
  \partial_nI_{\rm bulk}[\widehat B_n]\big|_{\partial T_\epsilon}
  &=
  \frac{1}{16\pi G_N}
  \int_0^{2\pi}d\tau\int du\,
  \frac{R}{n}\rho_h(n)\frac{2}{nR}
  \nonumber\\
  &=
  \frac{\rho_h(n)L_u}{4G_Nn^2}
  =
  \frac{A_h(n)}{4G_Nn^2}.
  \label{eq:dong-cbc-tube-area-new}
\end{align}
At \(n=1\), this gives \(A_h(1)/(4G_N)\), in agreement with
\eqref{eq:dong-cbc-area-law-new}.

\subsection{RT surface for Global AdS}
In previous sections, we saw that the entropy of a subregion is given by the invariant length of the brane divided by $4 G_N$. In this section, we are going to perform an explicit computation of the on-shell action of replicated geometry $B_n$ for a subregion in the vacuum state. The metric can still be written as
\begin{equation}
  ds_n^2
  =
  f_n(\rho)d\tau_E^2+\frac{d\rho^2}{f_n(\rho)}+\rho^2du^2,
  \qquad
  f_n(\rho)=\frac{\rho^2-\rho_h(n)^2}{\ell^2}.
  \label{eq:dong-cbc-adapted-metric-new}
\end{equation}
We have determined \(\rho_h(n)\) and
\(\rho_c(n)\) in terms of boundary data.
\begin{equation}
  \rho_h(n)
  =
  \frac{\ell}{2n}(k-\Delta),
  \qquad
  \rho_c(n)^2
  =
  \frac{\ell^2}{2n^2}\frac{k-\Delta}{\Delta}.
  \label{eq:dong-cbc-rhoh-rhoc-new}
\end{equation}
So far, this is a general computation that holds for any time-reversal-symmetric state with $U(1)$ and translational symmetry. For global AdS, 
we can map the above coordinates to global coordinates using the isometry of hyperbolic space. To find the entropy, we need $L_u$ as we already have $\rho_h$ for any set of boundary conditions.
For
physical \(n=1\) horizon segment we write
\begin{equation}
  C_1:\qquad u\in[0,u_{\max}],
  \qquad
  L_u=u_{\max}.
  \label{eq:umax-definition-new}
\end{equation}
The goal is to find $L_u$ in terms of the boundary subregion.\\

We specialize to a global interval \(A=[-\phi_0,\phi_0]\) on the
\(t_E=0\) slice of the physical CBC boundary at \(\varrho=\varrho_c\) \footnote{$\varrho$ is the radial coordinate for global AdS.}.  Euclidean
AdS\(_3\) is the hyperboloid \(H^3\subset\mathbb R^{1,3}\) with \emph{four}
embedding coordinates,
\begin{equation}
  -X_{-1}^2+X_0^2+X_1^2+X_2^2=-\ell^2 ,
  \label{eq:ads3-hyperboloid}
\end{equation}
The global chart is
\[
  X_{-1}=\sqrt{\ell^2+\varrho^2}\,\cosh\tfrac{\tau_E}{\ell},\quad
  X_0=\sqrt{\ell^2+\varrho^2}\,\sinh\tfrac{\tau_E}{\ell},\quad
  X_1=\varrho\cos\phi,\quad
  X_2=\varrho\sin\phi ,
\]
In these global coordinates, the metric for the AdS$_3$ can be written as
\begin{eqnarray}
 ds^2= (1+\varrho^2/\ell^2)d\tau_E^2+\frac{d\varrho^2}{(1+\varrho^2/\ell^2)}+\varrho^2 d\phi^2.
\end{eqnarray}
So on the time-symmetric slice \(\tau_E=0\) one has \(X_0=0\) and
\(X_{-1}=\sqrt{\ell^2+\varrho^2},\ X_1=\varrho\cos\phi,\ X_2=\varrho\sin\phi\).

The corresponding adapted (Rindler / hyperbolic-black-hole) chart of
\eqref{eq:dong-cbc-adapted-metric-new} uses all four embedding coordinates as
\begin{equation}
\begin{aligned}
  Y_{-1}&=\frac{\ell}{\rho_h(n)}\,\rho\cosh\!\left(a_n u\right),
  &\qquad
  Y_2&=\frac{\ell}{\rho_h(n)}\,\rho\sinh\!\left(a_n u\right),\\[2pt]
  Y_1&=\frac{\ell}{\rho_h(n)}\sqrt{\rho^2-\rho_h(n)^2}\,
        \cos\!\Big(\tfrac{\rho_h(n)}{\ell^2}\tau_E\Big),
  &\qquad
  Y_0&=\frac{\ell}{\rho_h(n)}\sqrt{\rho^2-\rho_h(n)^2}\,
        \sin\!\Big(\tfrac{\rho_h(n)}{\ell^2}\tau_E\Big),
\end{aligned}
  \label{eq:bulk-rindler-map-new}
\end{equation}
with $a_n\equiv\frac{\rho_h(n)}{\ell}.$
A direct substitution shows \eqref{eq:bulk-rindler-map-new} satisfies the
hyperboloid constraint \eqref{eq:ads3-hyperboloid} identically and reproduces
the full metric \eqref{eq:dong-cbc-adapted-metric-new}.\\

The induced metric
caps off smoothly at \(\rho=\rho_h\) with period
\(\beta_n=2\pi\ell^2/\rho_h\).  On the slice \(\tau_E=0\) the time-embedding
coordinate vanishes, \(Y_0=0\), and
\(Y_1=\frac{\ell}{\rho_h(n)}\sqrt{\rho^2-\rho_h(n)^2}\), reducing
\eqref{eq:ads3-hyperboloid} to the \(H^2\) form \(-Y_{-1}^2+Y_1^2+Y_2^2=-\ell^2\).\\

Explicitly, now we do a single \(H^3\) boost (an
\(SO(1,1)\subset SO(1,3)\) isometry) acting only in the \((Y_{-1},Y_1)\) plane, so that we have
\begin{equation}
  X_{-1}=\cosh\gamma\,Y_{-1}+\sinh\gamma\,Y_1,
  \quad
  X_1=\sinh\gamma\,Y_{-1}+\cosh\gamma\,Y_1,
  \quad
  X_2=Y_2,
  \quad
  X_0=Y_0.
  \label{eq:explicit-boost}
\end{equation}
The rapidity \(\gamma\) is fixed by demanding that the boosted brane endpoint land
on the interval endpoint \(P_+=(\sqrt{\ell^2+\varrho_c^2},\,
\varrho_c\cos\phi_0,\,\varrho_c\sin\phi_0)\):
\begin{equation}
  \tanh\gamma=\frac{\varrho_c\cos\phi_0}{\sqrt{\ell^2+\varrho_c^2}}
  \qquad\Longleftrightarrow\qquad
  \sinh\gamma=\frac{\varrho_c\cos\phi_0}{\sqrt{\ell^2+\varrho_c^2\sin^2\phi_0}}
  =\frac{\varrho_M}{\ell},
  \label{eq:boost-rapidity}
\end{equation}
where \(\varrho_M=\ell\sinh\gamma\) is the radius of the geodesic's turning point
at \(\phi=0\)\footnote{ The unboosted brane has \(Y_1=0\) at \(\tau_E=0\), so at \(u=0\)
it sits at the center \(\varrho=0\); the boost \eqref{eq:explicit-boost}
translates its turning point out to \(\varrho_M\) so that its two ends reach the
wall at \(\phi=\pm\phi_0\). For \(\phi_0\to\pi/2\), \(\gamma\to0\) and the
geodesic is the diameter through the centre; for small \(\phi_0\),
\(\varrho_M\to\varrho_c\) and the geodesic hugs the wall.}. Since
\eqref{eq:explicit-boost} gives \(X_2=Y_2\) (and \(X_0=Y_0=0\) on the slice),
at the horizon where \(\rho=\rho_h(n)\) and at \(\tau_E=0\), we have
\begin{equation}
  X_2
  =
  \ell\sinh\!\left(a_nu\right).
  \label{eq:horizon-X2-map-new}
\end{equation}
At the endpoint \(\phi=\phi_0\), the same embedding coordinate is
\(X_2=\varrho_c\sin\phi_0\).  Hence the bulk endpoint map gives, in the
\(n\)-th adapted coordinate convention,
\begin{equation}
  \sinh\!\left(a_nu_{\max}^{(n)}\right)
  =
  \frac{\varrho_c}{\ell}\sin\phi_0 .
  \label{eq:bulk-umax-relation-new}
\end{equation}
Equivalently,
\begin{equation}
  u_{\max}^{(n)}
  =
  \frac{\ell}{\rho_h(n)}
  \operatorname{arcsinh}
  \left[
  \frac{\varrho_c}{\ell}\sin\phi_0
  \right].
  \label{eq:bulk-umax-general-new}
\end{equation}
Using the global CBC boundary relations
\[
  \frac{\varrho_c^2}{\ell^2}
  =
  \frac{k-\Delta}{2\Delta}, \quad \rho_h(n)=\frac{\ell}{2n}(k-\Delta),
\]

this becomes
\begin{equation}
  u_{\max}^{(n)}
  =
  \frac{2n}{k-\Delta}
  \operatorname{arcsinh}
  \left[
  \sqrt{\frac{k-\Delta}{2\Delta}}\,
  \sin\phi_0
  \right].
  \label{eq:bulk-umax-k-new}
\end{equation}
The \(n=1\) value entering the entropy calculation is therefore
\begin{equation}
  u_{\max}
  \equiv
  u_{\max}^{(1)}
  =
  \frac{2}{k-\Delta}
  \operatorname{arcsinh}
  \left[
  \sqrt{\frac{k-\Delta}{2\Delta}}\,
  \sin\phi_0
  \right].
  \label{eq:bulk-umax-n1-new}
\end{equation}
For a global interval $A=[-\phi_0,\phi_0]$, there are two boundaries of the region. In our analysis, each one yields an identical contribution. 
Hence, the entropy of the region is given by 
 
\begin{equation}
  S_A
  =
  \frac{\ell}{2G_N}
  \operatorname{arcsinh}
  \left[
  \sqrt{\frac{k-\Delta}{2\Delta}}\,
  \sin\phi_0
  \right].
  \label{eq:bulk-entropy-umax-new}
\end{equation}
Because \(\rho_h(1)=\ell(k-\Delta)/2\),  the factor
\((k-\Delta)^{-1}\) in \(u_{\max}\) gets canceled from $\rho_h$.
Next, we directly compute the geodesic length and find agreement with the above analysis.
\subsection*{Subregion entropy for Global AdS}
\label{sec:global-geodesic-computation}
The metric of global AdS\(_3\) in global coordinates is
\begin{equation}
  ds^2_{t=0}
  =
  \frac{d\varrho^2}{1+\varrho^2/\ell^2}
  +\varrho^2 d\phi^2,
  \qquad
  \phi\sim\phi+2\pi .
  \label{eq:global-slice}
\end{equation}
The CBC boundary is a constant-\(\varrho\) surface whose trace of extrinsic curvature is held fixed.  With
\[
  k\equiv K\ell,
  \qquad
  \Delta\equiv\sqrt{k^2-4}.
\]
The extrinsic curvature is
\begin{equation}
  K
  =
  \frac{\ell^2+2\varrho_c^2}
  {\ell\,\varrho_c\sqrt{\ell^2+\varrho_c^2}},
  \qquad
  \frac{\varrho_c^2}{\ell^2}
  =
  \frac{k-\Delta_K}{2\Delta_K}.
  \label{eq:global-wall-K-data}
\end{equation}
This fixes the endpoint radius \(\varrho_c\). Now take the interval
\begin{equation}
  A=[-\phi_0,\phi_0],
  \qquad
  0<\phi_0<\pi ,
\end{equation}
on the \(t=0\) wall slice.  The bulk geodesic length is computed from the bulk
metric \eqref{eq:global-slice} with endpoints
\((\varrho,\phi)=(\varrho_c,\pm\phi_0)\).

For a curve \(X^\mu(\lambda)=(\varrho(\lambda),\phi(\lambda))\), the length
functional is
\begin{equation}
  L_{\rm geo}
  =
  \int d\lambda\,
  \sqrt{
  \frac{\dot\varrho^2}{1+\varrho^2/\ell^2}
  +\varrho^2\dot\phi^2}.
  \label{eq:global-geodesic-functional}
\end{equation}
Set
\(\varrho=\ell\sinh\eta\), for which
\begin{equation}
  ds^2_{t=0}
  =
  \ell^2\left(d\eta^2+\sinh^2\eta\,d\phi^2\right).
\end{equation}
The endpoints are
\[
  (\eta,\phi)=(\eta_c,\pm\phi_0),
  \qquad
  \sinh\eta_c=\frac{\varrho_c}{\ell}.
\]
The hyperbolic distance between these two points satisfies
\begin{align}
  \cosh\frac{L_{\rm geo}}{\ell}
  &=
  \cosh^2\eta_c-\sinh^2\eta_c\cos(2\phi_0)
  \nonumber\\
  &=
  1+2\frac{\varrho_c^2}{\ell^2}\sin^2\phi_0 .
\end{align}
Therefore
\begin{equation}
  L_{\rm geo}
  =
  2\ell\,
  \operatorname{arcsinh}
  \left[
  \frac{\varrho_c}{\ell}\sin\phi_0
  \right].
  \label{eq:direct-global-geodesic}
\end{equation}
Using \eqref{eq:global-wall-K-data}, this may be written as
\begin{equation}
  L_{\rm geo}
  =
  2\ell\,
  \operatorname{arcsinh}
  \left[
  \sqrt{\frac{k-\Delta}{2\Delta}}\,
  \sin\phi_0
  \right].
  \label{eq:global-geodesic-k}
\end{equation}
The corresponding geodesic entropy is
\begin{equation}
\boxed{
  S_{A,{\rm geo}}^{\rm CBC}(K,\phi_0)
  =
  \frac{L_{\rm geo}}{4G_N}
  =
  \frac{\ell}{2G_N}\,
  \operatorname{arcsinh}
  \left[
  \sqrt{\frac{K\ell-\sqrt{K^2\ell^2-4}}
  {2\sqrt{K^2\ell^2-4}}}\,
  \sin\phi_0
  \right].}
  \label{eq:cbc-entropy-global-phi0}
\end{equation}

\subsection{CHM's formalism} A complementary derivation, the Casini-Huerta-Myers (CHM) construction \cite{Casini:2011kv}, is a boundary field-theory argument, with a holographic realization. 
This derivation is often more transparent. It also directly connects with the preceding construction of the replicated geometry and the partition function $Z[n]$.
The dual boundary theory for gravity with conformal boundary conditions is conjectured to be the usual holographic CFT coupled to a time-like Liouville theory and deformed by a suitable marginal $T\bar{T}$ operator \cite{Allameh:2025gsa}. Here we sketch a brief derivation of applying the CHM formalism for calculating subregion entanglement entropy of the corresponding dual theory using holography.  

CHM provides a map that relates the reduced density matrix of a boundary spherical subregion in vacuum (state with no insertion) to a thermal density matrix on hyperbolic space. In other words, the entanglement entropy of the subregion is mapped to the thermodynamic entropy of this thermal state, and ultimately the answer is given by the thermal entropy of the AdS-Rindler horizon in units of 4$G_N$.


Next, we invoke AdS/CFT with a conformal boundary condition. The dual theory now lives at the finite radial hypersurface $r_c$. It is also equivalent to the surface with extrinsic curvature $ K \ell \geq 2$. The difference $K \ell -2,$ can be thought of as a measure of $r_c$, explicitly $K \ell -2 \propto 1/r_c$. But at this surface, one can evaluate the $u_{\rm max}$ and it is given by \eqref{eq:bulk-umax-n1-new}. Notice the difference in the cylinder calculation here. Now the $u_{\rm max}$ depends on the extrinsic curvature. This is a consequence of AdS isometry. This point will become highly relevant in the next section.


The gravity dual of a CFT on hyperbolic space is a hyperbolic black hole. The CFT has a temperature $T=\frac{1}{2 \pi R}$. The conformal boundary in gravity amounts to putting these boundary conditions for a hyperbolic black hole (see \cite{Casini:2011kv} for the explicit bulk coordinate transformation). We have already constructed such replicated geometry with index $n$ in the previous section \ref{sec:dong-argument-cbc}. The hyperbolic black-hole metric can be written as
\[
   ds^{2}=f(\rho)\,d\tau_{E}^{2}+\frac{d\rho^{2}}{f(\rho)}
            +\rho^{2}\,du^{2},\quad
   f(\rho)=\frac{\rho^{2}-\rho_{h}^{2}}{\ell^{2}}.
\]
with $\rho\in[\rho_{h},\rho_{c}]$,
$\tau_{E}\in[0,\beta]$, $u\in[0,u_{\max}]$.
To find the entropy, we follow the LM prescription and replicate the manifold with replica index $n$, and then we have $\rho\in[\rho_{h}(n),\rho_{c}(n)]$,
$\tau_{E}\in[0,\beta(n)]$ and $u\in[0,u_{\max}]$. We find the Rényi entropy $S_n$, and then the Von-Neumann entropy can be evaluated as \footnote{For a global interval $A=[-\phi_0,\phi_0]$, there are two boundaries of the region. In our analysis, each one yields an identical contribution. }
\begin{equation}
  S_A
  =
  \frac{\ell}{2G_N}
  \operatorname{arcsinh}
  \left[
  \sqrt{\frac{k-\Delta}{2\Delta}}\,
  \sin\phi_0
  \right].
  \label{eq:bulk-entropy-umax-new}
\end{equation}
\section{Explicit calculations}
We have established the RT formula for the gravity theory having a conformal boundary condition. This amounts to finding the length of the minimal geodesic in AdS bulk and anchored at the subregion end point. Using this formalism we are going to work out an example of (non)rotating BTZ black hole. Furthermore, we write the entropy in terms of boundary data which is the extrinsic curvature and the conformal temperature (see \eqref{nonroatingbtzgeodesic} and \eqref{rotatingbtzgeodesic}).
\subsection{Subregion entropy for Non-rotating BTZ}
In this section, we use the RT formula to calculate the entanglement entropy.
\begin{eqnarray}
    S_A = \frac{\mathrm{Area(min\,\,\, surface)}}{4 G_N}.
    \label{RT formula}
\end{eqnarray}
Specifically, we start with a non-rotating BTZ black hole in the Lorentzian signature.
\begin{equation}
    ds^2 = -(r^2-m)dt^2 + \frac{dr^2}{(r^2-m)} + r^2dx^2.
\end{equation}
Following along the same lines as HRT \cite{Hubeny:2007xt}, we pick the subregion $\mathcal{A}$ on the boundary with coordinates $(t,x)$. At fixed time, the coordinates of the endpoints of $\mathcal{A}$ are $(r_c,t_0,-h)$ and $(r_c,t_0,h)$. Hence, the length of the interval is $l=2h$ on the conformal boundary. Here, the boundary of the spacetime is at a finite radius $r_c$. We solve for the RT surface $r(x)$ at this fixed time $t_0$. The Killing direction $x$ gives the following conservation:
\begin{equation}
    \frac{dr}{dx}=r\sqrt{(r^2-m)\left(\frac{r^2}{r_*^2}-1\right)},
\end{equation}
where we defined $r_*$ as the turning point of the geodesic which is in the interior and at $x=0$. This gives the relation between $x$ and $r$ as
\begin{equation}
     x = -\frac{1}{2 \sqrt{m}}\log \left( \frac{-2r_* \sqrt{m(r^2-m)(r^2-r_*^2)-2mr_*^2 + r^2r_*^2 +mr^2}}{r^2(r_*^2-m)}\right).
\end{equation}
Since the endpoints are at $x=\pm h$ and $r=r_c$,
\begin{equation}
    h = -\frac{1}{2 \sqrt{m}}\log \left( \frac{-2r_* \sqrt{m(r_c^2-m)(r_c^2-r_*^2)-2mr_*^2 + r_c^2r_*^2 +mr_c^2}}{r_c^2(r_*^2-m)}\right).
\end{equation}
Inverting this gives 
\begin{equation}
    r_* = \frac{\sqrt{m} r_c \left(e^{2 h \sqrt{m}}+1\right)}{\sqrt{-2 r_c^2 e^{2 h \sqrt{m}}+r_c^2 e^{4 h \sqrt{m}}+4 m e^{2 h \sqrt{m}}+r_c^2}}.
\end{equation}
Now the length of the geodesic is
\begin{equation}
\begin{split}
    L &= 2\int_{r_*}^{r_c} \frac{rdr}{r_* \sqrt{(r^2-m)\left( \frac{r^2}{r_*^2}-1\right)}} \\
    & = - 2 \log \frac{\sqrt{r_c^2-m}-\sqrt{r_c^2-r_*^2}}{\sqrt{r_*^2-m}}.
\end{split}
\end{equation}
Substituting $r_*$ in $L$ and using \eqref{RT formula} gives \\
\begin{equation}
\label{nonrotbtz}
    S = -\frac{1}{2 G_N}\log \frac{\sqrt{r_c^2 \sinh^2{(h\sqrt{m})}+m}-r_c\sinh{(h\sqrt{m})}}{\sqrt{m}}.
\end{equation}
In the conformal case, we trade the cutoff surface $r=r_c$ in favor of extrinsic curvature $K$.
The trace of extrinsic curvature at the cutoff surface at constant $r=r_c$ is:
\begin{eqnarray}\label{eq:K_rc}
    K = \frac{2r_c^2-m}{r_c\sqrt{r_c^2-m}}.
\end{eqnarray}
And we need to invert the above relation to find $r_c$ in terms of $K$.
The above equation can be written in terms of inverse temperature (not the conformal temperature yet) as
\begin{equation}
S=\frac{1}{2G_N}\log\!\left[
\sqrt{1+\left(\frac{\beta r_c}{2\pi}\right)^2
\sinh^2\!\left(\frac{2\pi h}{\beta}\right)}
+\frac{\beta r_c}{2\pi}\sinh\!\left(\frac{2\pi h}{\beta}\right)
\right].
\label{subregion entropy BTZ}
\end{equation}
In the large $r_c$ limit we recover, as expected, the Dirichlet case, 
\begin{equation}
    S=\frac{1}{2G}\log\left(\frac{r_c\,\beta}{\pi}\sinh\left(\frac{2\pi h}{\beta}\right)\right).
\end{equation}
where $\beta=2\pi/\sqrt{m}, 2h=L_A$ and $r_\infty=1/\epsilon$, which is in perfect agreement with HRT.

Next, we define 
\[
x\equiv \frac{\beta r_c}{2\pi}=\frac{r_c}{\sqrt m},
\qquad
L_A\equiv 2h,
\qquad
\Delta\equiv \sqrt{K^2\ell^2-4},
\]
where we have introduced explicit factors of the AdS curvature scale $\ell$. With these definitions, the  entropy \eqref{subregion entropy BTZ} is,
\begin{equation}
  S(K,L_A,\beta)
=
\frac{\ell}{2G_N}\log\!\left[
\sqrt{1+\frac{K\ell+\Delta}{2\Delta}\,
\sinh^2\!\left(\frac{\pi L_A}{\beta}\right)}
+\sqrt{\frac{K\ell+\Delta}{2\Delta}}\,
\sinh\!\left(\frac{\pi L_A}{\beta}\right)
\right].  
\end{equation}

\noindent We can now use the conformal-temperature relation for non-rotating BTZ:
\[
\tilde\beta
=
\beta\,\frac{\sqrt{r_c^2-m}}{\ell r_c}
=
\beta\,\frac{\sqrt{x^2-1}}{\ell x}.
\]
Substituting this into \(S(K,L_A,\beta)\), we obtain
\begin{align}
\label{nonroatingbtzgeodesic}
    S(K,L_A,\tilde\beta)=
\frac{\ell}{2G_N}\log\!\left[
\sqrt{1+\frac{K\ell+\Delta}{2\Delta}\,
\sinh^2\!\left(\frac{\pi (K\ell-\Delta)}{2\tilde\beta}\,L_A\right)}
+\sqrt{\frac{K\ell+\Delta}{2\Delta}}\,
\sinh\!\left(\frac{\pi (K\ell-\Delta)}{2\tilde\beta}\,L_A\right)
\right].\nonumber\\
\end{align}
It can be compactly written as
\begin{equation}
\boxed{
    S(K,L_A,\tilde{\beta})= \frac{\ell}{2G_N}\,
\operatorname{arcsinh}\!\left[
\sqrt{\frac{K\ell+\Delta}{2\Delta}}\,
\sinh\!\left(
\frac{\pi (K\ell-\Delta)}{2\tilde\beta}\,L_A
\right)
\right].}
\end{equation}
For a large interval limit or a high temperature limit, we have
\begin{eqnarray}
   S(K,L_A,\tilde\beta)
\sim
\frac{\pi \ell (K \ell-\Delta)}{4G_N\,\tilde\beta}\,L_A
+\frac{\ell}{4G_N}\log\!\left(\frac{K \ell+\Delta}{2\Delta}\right).
\end{eqnarray}

The leading contribution can be directly compared with the standard Cardy entropy by identifying the effective central charge, expressed in terms of the matter central charge
$c_{\rm eff}=\frac{3 \ell}{4G_N}\left(K \ell-\Delta\right)$,
\begin{equation}
S_{\rm leading}(L_A,\tilde\beta)
=
\frac{\pi c_{\rm eff}}{3\tilde\beta}\,L_A.
\end{equation}
This is in perfect agreement with the derivation done in Allameh et.al. \cite{Allameh:2025gsa}. All the details of the calculations in this subsection can be found in  Appendix \ref{calculation}. \\

\noindent \textbf{Massless BTZ case}\\
In a boundary theory with vanishing anomaly central charge, the state prepared without any insertions maps to a state with zero energy in the bulk, which corresponds to the $m=0$ BTZ solution.
The zero temperature ($m=0$) limit of the BTZ black hole has the following geometry:
\begin{equation}
    ds^2=-r^2 dt^2 + \frac{dr^2}{r^2}+r^2 dx^2.
\end{equation}
The geodesic satisfies:
\begin{equation}
   \frac{dr}{dx}= r^2\sqrt{\frac{r^2}{r_*^2}-1}. 
\end{equation}
Again, defining the subregion at $r=r_c$ and $x=\pm h$ gives the turning point:
\begin{equation}
    r_*=\frac{r_c}{\sqrt{1+h^2r_c^2}}.
\end{equation}
So the geodesic length, for any finite $r_c$, is:
\begin{equation}
    L = 2 \log \left( \frac{r_c + \sqrt{r_c^2-r_*^2}}{r_*} \right)=2\, \text{arcsinh} \, (h r_c).
\end{equation}
From  \eqref{eq:K_rc}, for $m=0$ we obtain $K \ell = 2$.  Thus, we cannot invert $r_c$ in terms of $K$. 
Note that this case is degenerate since surfaces with different $r_c$ have the same $K \ell=2$. If we take the large $r_c$ limit and write $r_c$ in terms of the cutoff $r_c \sim \frac{1}{\epsilon}$, we have,
\begin{equation}
\label{masslessbtz}
    S = \frac{1}{2 G_N}\log \left( \frac{2 h}{\epsilon} \right)=\frac{c_m}{3}\log \left( \frac{2 h}{\epsilon} \right).
\end{equation}
Thus, the entanglement entropy can be written in terms of the boundary UV cutoff. 
\eqref{masslessbtz} is consistent with the subregion entropy in Poincaré AdS$_3$.






\subsection{Subregion entropy for Rotating BTZ}

We now turn to the rotating BTZ metric with a conformal boundary:
\begin{equation}
    ds^2=-\frac{(r^2-r_+^2)(r^2-r_-^2)}{r^2}dt^2+\frac{r^2}{(r^2-r_+^2)(r^2-r_-^2)}dr^2+r^2(dx+\frac{r_+r_-}{r^2}dt)^2\nonumber.
\end{equation}
At \(t=0\), for an interval of coordinate length \(L_A\), the geodesic length in rotating BTZ gives (more detailed analysis is done in Appendix \ref{rotating_cal})
\[
\cosh L_{\rm geo}
=
\frac{r_c^2-r_-^2}{r_+^2-r_-^2}\cosh(r_+ L_A)
-
\frac{r_c^2-r_+^2}{r_+^2-r_-^2}\cosh(r_- L_A).
\]
Hence, the RT formula yields the entropy as
\begin{equation}
S_{\rm rot}
=
\frac{L_{\rm geo}}{4G}
=
\frac{1}{4G}
\operatorname{arccosh}\!\left[
\frac{r_c^2-r_-^2}{r_+^2-r_-^2}\cosh(r_+ L_A)
-
\frac{r_c^2-r_+^2}{r_+^2-r_-^2}\cosh(r_- L_A)
\right].
\end{equation}
Now we use the conformal quantity and write the EE as
\begin{align}
\label{rotatingbtzgeodesic}
\boxed{
S_{\rm rot}(K,\tilde\beta,\tilde\Omega,L_A)
=
\frac{\ell}{4G}
\operatorname{arccosh}\!\left[
\frac{K\ell+\Delta}{2\Delta}\,
\cosh\!\left(
\frac{\pi (K\ell-\Delta)}{\tilde\beta(1-\tilde\Omega^2)}\,L_A
\right)
-
\frac{K\ell-\Delta}{2\Delta}\,
\cosh\!\left(
\frac{2\pi \tilde\Omega}{\tilde\beta(1-\tilde\Omega^2)}\,L_A
\right)
\right],}
\end{align}
with $
\Delta=\sqrt{K^2\ell^2-4}.$

At high conformal temperature, we have
\begin{equation}
S_{\rm rot}(K,\tilde\beta,\tilde\Omega,L_A)
\sim
\frac{\pi\ell (K\ell-\Delta)}{4G\,\tilde\beta(1-\tilde\Omega^2)}\,L_A
+
\frac{\ell}{4G}\log\!\left(\frac{K\ell+\Delta}{2\Delta}\right).
\end{equation}
Using
\[
c_{\rm eff}=\frac{3\ell}{4G}(K-\Delta),
\]
the leading term becomes
\[
S_{\rm leading}
=
\frac{\pi c_{\rm eff}}{3\,\tilde\beta(1-\tilde\Omega^2)}\,L_A.
\]
This is in agreement with the Allameh et.al. \cite{Allameh:2025gsa} for Cardy entropy.


\section{Entropy from the dual boundary theory}
\label{sec:chm_boundary}

The entanglement entropy of a CFT deformed by the $T\bar{T}$ operator has been calculated in \cite{Chen:2018eqk,Donnelly:2018bef,Lewkowycz:2019xse}. In this section, we discuss the subregion entanglement entropy of our boundary theory (which consists of the matter CFT coupled to the timelike Liouville theory and deformed by the $T\bar{T}$) on a cylinder in the vacuum state, i.e., a state with no insertion, which is not the ground state.\\

\textbf{An aside:-}
By ``ground state,'' we mean the state with minimum energy. This is different from the identity state obtained by inserting the identity operator at the origin in radial quantization. In an ordinary unitary CFT, the two states coincide because the identity is the lowest-energy state. In the present non-unitary theory, however, they need not be the same. The spectrum contains a lowest-weight sector satisfying
\begin{equation}
    h_{\min}=\bar h_{\min}<0,
\end{equation}
and the long Euclidean evolution projects onto this sector. Since
\begin{equation}
    c_{\mathrm{tot}}=0,
    \qquad
    c_{\mathrm{eff}}
    =
    c_{\mathrm{tot}}-24h_{\min}
    =
    -24h_{\min},
\end{equation}
Its cylinder energy is
\begin{equation}
    E_0
    =
    \frac{h_{\min}+\bar h_{\min}-c_{\mathrm{tot}}/12}{R}
    =
    -\frac{c_{\mathrm{eff}}}{12R},
    \qquad
    \widetilde E_0\equiv R E_0
    =
    -\frac{c_{\mathrm{eff}}}{12}.
\end{equation}

The CHM construction can be applied to any state in the CFT. With this insertion, the reduced density matrix involves computing the 2n-point correlators. We will return to this in the future. The corresponding bulk state is the minimal-energy global AdS$_3$ solution. The \(M=0\) BTZ geometry instead corresponds to the zero-energy identity sector, which is also somewhat degenerate, as discussed before.

Using the CHM map, we can embed the region's domain of dependence into hyperbolic space as $\mathbb{R}\times H^1$. In Euclidean signature, it is a thermal cylinder ( with $u \in [0,u_{max}]$) with $\tau \sim \tau+\beta$. The temperature can be written as $T=1/2 \pi R$, where $R$ is the size of the region. The EE can be calculated using the twist operator, but in our deformed theory coupled to Liouville fields, it is non-trivial to construct the twist operator. Hence, we follow the same strategy that we did in the bulk. We will find $Z(n)$, and using the derivative with respect to n, we will find the entropy. Calculating entropy involves several steps.
\begin{enumerate}
    \item We set up the replicated geometry and compute the matter--CFT
partition function and thermal stress tensor on
$\mathbb{R}_\tau\times H^1_u$. 
\item Write the
Liouville saddle equation and solve the  saddle equation. First, we find the solution in leading order in coupling and later we find all order result.
\item We evaluate the corresponding on-shell action and take replica index n derivative and find the entropy.
\end{enumerate}

\subsection*{Replicated geometry on $\mathbb{R}\times H^1$}
\label{chm:sec-geom}
To arrive at the replicated geometry for $Z(n)$, we employ the CHM transformation \cite{Casini:2011kv} directly on the boundary metric.
Start from the Lorentzian cylinder
$ds^{2}=-dt^{2}+R^{2}d\phi^{2}$ with the CFT in its vacuum and the
entangling interval $A=[-\phi_0,\phi_0]$ on $t=0$.
\begin{eqnarray}
    ds^2= -dt^2+ R^2 d \phi^2.
\end{eqnarray}
The entangling surface is a point\footnote{ The entangling surface could be a union of points. An interval on the cylinder has two entangling points. For each entangling point, the entropies are essentially the same, hence we multiply the entropy calculated at $\phi=\phi_0$ by 2. } at $\phi= \phi_0$ on the $t=0$ surface. We do the following coordinate transformation (CHM map \cite{Casini:2011kv})
\begin{eqnarray}
    \tan (t/R)= \frac{\sin \phi_0\, \sinh (\tau/R)}{\cosh u +\cos \phi_0 \cosh (\tau/R)}\nonumber,\\
    \tan \phi= \frac{\sin \phi_0\, \sinh u}{\cos \phi_0 \cosh u+\cosh(\tau/R)}.
    \label{CHM map}
\end{eqnarray}
Then the metric on the cylinder becomes
\begin{gather}
    ds^2= \Omega^2 (-d\tau^2+R^2 du^2), \\
    \Omega^2= \frac{\sin^2 \phi_0}{(\cosh u+\cos \phi_0 \cosh (\tau/R))^2+\sin^2 \phi_0\, \sinh^2(\tau/R)}.
\end{gather}

One can perform a conformal transformation to eliminate the prefactor $\Omega^2$, and the resulting metric on the boundary (now hyperbolic space $\mathcal{H}$) is of the form $ \mathbb{R}\times H^1 \cong \mathbb{R}\times \mathbb{R} $,
\begin{eqnarray}
    ds^2= -d\tau^2+R^2 du^2.
\end{eqnarray}
We can see the entangling point and its causal domain of dependence can be mapped as 
\begin{eqnarray}
   &&  (t,\phi)= (\pm R \phi_0,0), \quad \tau \rightarrow \pm \infty\nonumber\\
  &&  (t,\phi) = (0,\phi_0), \quad  u \rightarrow \infty,\tau \rightarrow 0.
\end{eqnarray}
The new coordinates $\tau, u$ precisely cover the causal domain of dependence of the entangling region. The vacuum state on the cylinder (or entangling region) is now mapped to the thermal states on $\mathcal{H}$ with temperature $T=1/2 \pi R$,
and the resulting metric (now hyperbolic space $\mathcal{H}$) is of the form $ \mathbb{R}\times H^1 \cong \mathbb{R}\times \mathbb{R}_u $:
\begin{eqnarray}
    ds^2= -d\tau^2+R^2 du^2= R^2( - d \tilde{\tau}^2+du^2).
\end{eqnarray}
The conformal temperature is defined by the periodicity of $\tilde{\tau}$ and it is $\tilde{\beta}= 2 \pi$. When we replicate the manifold, we have $\tilde{\beta}_n= 2 \pi n$. The relevant partition function can be written as the trace of the Hilbert space as
\begin{equation}
    Z(\tilde{\beta})= \mathrm{Tr\,\,exp}[-\tilde{\beta}\tilde{H}].
\end{equation}
Here $\tilde{H}\equiv \lambda H^{CBC}$ is the generator of time translation of $\tilde{\tau}$ and $H^{CBC}$ is the Hamiltonian with conformal boundary condition.

We can also map the cutoff ($t=0$ slice $\rightarrow \tau =0$ slice) on the cylinder to the hyperbolic space as 
\begin{eqnarray}
    \tan (\phi_0-\delta \phi)=\frac{\sin \phi_0 \sinh u_{max}}{\cos \phi_0 \cosh u_{max}+1}.
\end{eqnarray}
On solving for $u_{max}$, we have
\begin{align}
u_{\max}
&=
2\,\operatorname{arctanh}\!\left(
\frac{
\tan\!\left(\frac{\phi_0-\delta\phi}{2}\right)
}{
\tan\!\left(\frac{\phi_0}{2}\right)
}
\right)
=
\ln\!\left[
\frac{
1+\dfrac{\tan\!\left(\frac{\phi_0-\delta\phi}{2}\right)}
        {\tan\!\left(\frac{\phi_0}{2}\right)}
}{
1-\dfrac{\tan\!\left(\frac{\phi_0-\delta\phi}{2}\right)}
        {\tan\!\left(\frac{\phi_0}{2}\right)}
}
\right],
\label{critical}
\\
u_{\max}
&\simeq
\ln\!\left(\frac{2\sin\phi_0}{\delta\phi}\right)
\qquad
(\delta\phi\ll 1).
\nonumber
\end{align}


Some comments are in order: This $u_{\rm max}$ is obtained when we consider the CFT living on a cylinder of radius $R$. This is a consequence of the conformal transformation and doesn't depend on the details of the theory, such as the Liouville coupling $\mu$. There is no connection to the bulk AdS spacetime here. \\

After a Weyl rescaling by $\Omega^{-2}$ the
\emph{fiducial} metric is
\begin{equation}
\label{chm:fid-Lor}
   d\tilde s^{2}=-d\tau^{2}+R^{2}du^{2}.
\end{equation}
Now we Wick rotate $\tau\to\tau_{E}$, and on the cylinder we have a thermal state at $T=1/(2\pi R)$. For the replica index $n$, the thermal circle has period
\begin{equation}
\label{chm:beta-dimensionful}
   \beta_n=2\pi R n .
\end{equation}
Introduce the dimensionless Euclidean time
$\tilde\tau\equiv\tau_{E}/R$ and denote its replicated period by
$\tilde\beta_n$ (also called the conformal temperature),
\begin{equation}
\label{chm:fid-conformal}
   d\tilde s^{2}_{\rm conf}=d\tilde\tau^{2}+du^{2},\qquad
   \tilde\tau\sim\tilde\tau+\tilde\beta_n,\qquad
   \tilde\beta_n\equiv\frac{\beta_n}{R}=2\pi n,\qquad
   u\in[0,u_{\max}].
\end{equation}
The geometry is a \emph{flat cylinder }- : $\tilde R=0$, no angular twist
($\tilde\Omega=0$), Euler characteristic $\chi=0$. 



The fiducial volume on the replica is
\begin{equation}
\label{chm:Vol}
   V_{n}\equiv\int d^{2}x\sqrt{\tilde g}\;
   =\tilde\beta_{n}\cdot u_{\max}
   =2\pi n\,u_{\max}.
\end{equation}

\subsection*{Deformed boundary-theory action}
The dual theory is given by a timelike Liouville mode modified by the deformation \footnote{Here the Liouville field $\Phi$ is the transformed field after the CHM map. In the appendix \ref{app:chm frame consistency} we discuss the invariance of Liouville action under CHM map.}
\begin{equation}
    \frac{\partial S}{\partial\lambda}=\int d^2x \sqrt{\tilde{g}}\tilde{T}\tilde{\bar{T}}e^{-2\Phi} +\frac{c_m}{48\pi\lambda}\int d^2x\sqrt{\tilde{g}}\tilde{R}.
\end{equation}
At linear order in the deformation, the action is given by:
\begin{equation}
\label{chm:S}
   S[\Phi,\tilde g]
   =S_{m}[\tilde g]
   +\frac{1}{4\pi b^{2}}\!\int\! d^{2}x\sqrt{\tilde g}\Big[
     -(\tilde\nabla\Phi)^{2}-\tilde R\,\Phi
     +4\pi b^{2}\mu e^{2\Phi}
     +4\pi b^{2}\lambda\,\tilde T\tilde{\bar T}\,e^{-2\Phi}\Big].
\end{equation}
with $b^{2}=6/c_{m}$, $\mu=(K\ell-2)/(16\pi G\ell)$,
$\lambda=16\pi G\ell$, $c_{m}=3\ell/(2G)$, $\chi=0$. Here, the $\mu\lambda$-independent piece already
absorbed into $S_{m}$. With the fiducial metric of 
\eqref{chm:fid-conformal} we have $\tilde R=0$ which kills the linear-in-$\Phi$
term and the action reduces to
\begin{equation}
\label{chm:S-flat}
   S=S_{m}[\tilde g]
   +\!\int\! d^{2}x\sqrt{\tilde g}\Big[
       \mu e^{2\Phi}+\lambda\,\tilde T\tilde{\bar T}e^{-2\Phi}
       -\tfrac{1}{4\pi b^{2}}(\tilde\nabla\Phi)^{2}\Big].
\end{equation}
\



For a 2d CFT at inverse temperature $\tilde\beta$ on a non-compact
spatial line $[0,u_{\max}]$, the free-energy density is \cite{Bloete:1986qm,Affleck:1986bv}: 

\begin{equation}
\label{chm:fmatter}
   f_{m}=-\frac{\pi c_{m}}{6\,\tilde\beta^{2}}.
\end{equation}
The free energy and partition function on our cylinder are
\begin{align}
   F_{m}(n)&=f_{m}\cdot u_{\max}
            =-\frac{\pi c_{m}u_{\max}}{6\,\tilde\beta_{n}^{2}}
            =-\frac{c_{m}u_{\max}}{24\pi n^{2}},\\
  \log Z_{m}(n)&=-\tilde\beta_{n} F_{m}(n)
                 =\frac{c_{m}u_{\max}}{12 n}.
\end{align}
This is the  universal (extensive) part of the thermal partition function
of any 2d CFT on the non-compact line at $\tilde\beta_{n}=2\pi n$
regulated to length $u_{\max}$.

For a 2d CFT in this thermal state, holomorphic factorization gives
constant expectation values 
\begin{equation}
\label{chm:T-thermal}
   \langle \tilde{T}_{\tilde\tau\tilde\tau}\rangle
        =-\frac{\pi c_{m}}{6\tilde\beta_{n}^{2}},
   \qquad
   \langle \tilde{T}_{uu}\rangle=+\frac{\pi c_{m}}{6\tilde\beta_{n}^{2}},
   \qquad
   \langle \tilde{T}^{\mu}{}_{\mu}\rangle=0.
\end{equation}

The  $\tilde{T}\tilde{\bar{T}}$ operator can be identified with
\begin{equation}
\label{chm:TTbar}
   \mathcal T\equiv\tilde T\tilde{\bar T}\big|_{\rm matter}=\frac{1}{8}(\tilde{T}_{\mu\nu} \tilde{\bar{T}}^{\mu\nu}- (\tilde{T}^{\mu}_{\mu})^2)
   =\frac{\pi^{2}c_{m}^{2}}{144\,\tilde\beta_{n}^{4}}
   =\frac{c_{m}^{2}}{2304\,\pi^{2}n^{4}}\,.
\end{equation}
$\mathcal T$ is intensive: it does not depend on $u_{\max}$. Note that the deformation is from the stress tensor of the full theory including matter with Liouville kinetic terms and excluding the cosmological constant term ($\mu e^{2\Phi}$). But we work in a constant $\Phi$ ansatz, hence the Liouville kinetic parts will drop out, and we just proceed with the matter stress tensor for our calculation.

\subsection*{First-order saddle in $\lambda$}
\label{chm:sec-eom}

Varying \eqref{chm:S-flat} with respect to $\Phi$,
\begin{equation}
\label{chm:dSdPhi}
   \frac{\delta S}{\delta\Phi}
   =\frac{1}{4\pi b^{2}}\bigl(2\tilde\Box\Phi\bigr)
   +2\mu e^{2\Phi}-2\lambda\,\tilde T\tilde{\bar T}(\Phi)\,e^{-2\Phi}=0,
\end{equation}
i.e.
\begin{equation}
\label{chm:EOM}
   2\,\tilde\Box\Phi
   +8\pi b^{2}\!\left[\mu e^{2\Phi}
                -\lambda\,\tilde T\tilde{\bar T}(\Phi)\,e^{-2\Phi}\right]=0.
\end{equation}
This is the $\tilde R=0$ specialisation of
\cite{Allameh:2025gsa}~eq.~(3.18).

On the flat cylinder with translation invariance in $\tilde\tau$ and $u$
(in the bulk of the cylinder), We can have  a constant saddle
$\Phi=\Phi_{*}(n)$. Then $\tilde\Box\Phi=0$ and
\eqref{chm:EOM} reduces to the \emph{algebraic} eq.
\begin{equation}
\label{chm:saddle-algebraic}
   \mu\,e^{2\Phi_{*}}
   =\lambda\,\tilde T\tilde{\bar T}(\Phi_{*})\,e^{-2\Phi_{*}}.
\end{equation}


At leading order in $\lambda$, the operator $\tilde T\tilde{\bar T}$
is the thermal CFT value \eqref{chm:TTbar}: 

\begin{equation}
\label{chm:saddle-1st}
   \mu e^{2\Phi_{*}^{(1)}}=\lambda\mathcal T\,e^{-2\Phi_{*}^{(1)}}
   \;\Longrightarrow\;
   e^{4\Phi_{*}^{(1)}}=\frac{\lambda\mathcal T}{\mu}.
\end{equation}
Plug \eqref{chm:TTbar} and $\lambda\mu=K\ell-2$, we have
\begin{align}
\label{chm:saddle-1st-K}
   e^{4\Phi_{*}^{(1)}}
   &=\frac{\lambda^{2}\,\pi^{2}c_{m}^{2}}
          {144\,\tilde\beta_{n}^{4}\,\lambda\mu}
   =\frac{\lambda^{2}\,\pi^{2}c_{m}^{2}}
          {144\,(2\pi n)^{4}\,(K\ell-2)}.
\end{align}
Using $\lambda c_{m}=24\pi\ell^{2}$
($\lambda=16\pi G\ell$, $c_{m}=3\ell/(2G)$, so
$\lambda c_{m}=16\pi G\ell\cdot 3\ell/(2G)=24\pi\ell^{2}$),
\begin{equation}
\label{chm:saddle-1st-final}
   e^{4\Phi_{*}^{(1)}}
   =\frac{(24\pi\ell^{2})^{2}\pi^2}{144\,(2\pi n)^{4}\,(K\ell-2)}
   =\frac{\ell^{4}}{4\,n^{4}\,(K\ell-2)},
   \qquad
   e^{2\Phi_{*}^{(1)}}=\frac{\ell^{2}}{2 n^{2}\sqrt{K\ell-2}}.
\end{equation}
This is a saddle for the Liouville field in the linear order in $\lambda$, but now we can generalize this to all orders in $\lambda$.

\subsection*{All-orders $\tilde{T}\tilde{\bar{T}}$ operator and Liouville saddle}
In this section, we study all orders $\tilde{T}\tilde{\bar{T}}$ operator and derive the Liouville saddle. The field theory lives on the thermal cylinder
$S^{1}_{\tilde\beta_{n}}\times\mathbb R_{u}$, with the spatial line
regulated only at the end by $L=u_{\max}$. For homogeneous/constant $\Phi$, the
dressing replaces the fiducial $\tilde{T}\tilde{\bar{T}}$ coupling by
\begin{equation}
\label{chm:alpha-def}
   \alpha\equiv\lambda e^{-2\Phi}.
\end{equation}
On a cylinder, the deformed thermal free-energy density follows from the
homogeneous $T\bar T$ flow equation.
Writing
$W(\tilde{\beta_n},\alpha)\equiv L^{-1}\log Z_{\rm cyl}$, \footnote{The cylinder flow equation follows directly from the definition of the
deformation and from stress-tensor factorization in a homogeneous thermal
state.  We define
\[
   W(\tilde\beta,\alpha)\equiv {1\over L}\log Z_{\rm cyl}(\tilde\beta,L,\alpha),
   \qquad
   S_\alpha=S_{\rm CFT}
      +\alpha\int d^{2}x\,\mathcal T ,
\]
with
\[
   \mathcal T
   \equiv
   \tilde{T}_{zz}\tilde{T}_{\bar z\bar z}-\tilde{T}_{z\bar z}^{2}.
\]
Then
\[
   \partial_\alpha W
   =
   -\tilde\beta\,\langle \mathcal T\rangle_\alpha .
\]
For a zero-momentum homogeneous thermal state on the cylinder,
\[
   \langle T_{\tilde\tau\tilde\tau}\rangle
      =\partial_{\tilde\beta}W,
   \qquad
   \langle T_{uu}\rangle
      ={W\over \tilde\beta},
   \qquad
   \langle T_{\tilde\tau u}\rangle=0 .
\]
In the complex-coordinate convention \(z=u+i\tilde\tau\), one has
\[
   \mathcal T
   =
   \tilde{T}_{zz}\tilde{T}_{\bar z\bar z}-\tilde{T}_{z\bar z}^{2}
   =
   -{1\over 4}\,
   \tilde{T}_{\tilde\tau\tilde\tau}\tilde{T}_{uu}
\]
for a diagonal homogeneous stress tensor. Hence
\[
   \langle \mathcal T\rangle_\alpha
   =
   -{1\over 4}
   \left(\partial_{\tilde\beta}W\right)
   \left({W\over \tilde\beta}\right).
\]
Substituting this into the deformation equation gives the cylinder
Burgers equation
\begin{equation}
\label{chm:cyl-burgers}
   \partial_{\alpha}W
   =
   {1\over 4}\,
   W\,\partial_{\tilde\beta}W,
   \qquad
   W(\tilde\beta,0)
   =
   {\pi c_m\over 6\tilde\beta}.
\end{equation}
In the replica geometry we later set \(\tilde\beta=\tilde\beta_n=2\pi n\).}
\begin{equation}
\label{chm:cyl-burgers}
   \partial_{\alpha}W
   =\frac{1}{4}\,W\,\partial_{\tilde{\beta_n}}W.
   \qquad
 .
\end{equation}
Solving this first-order equation with boundary condition  $W(\tilde{\beta}_n,0)=\frac{\pi c_{m}}{6\tilde{\beta_n}}$ 
gives
\begin{equation}
\label{chm:cyl-logZ-alpha}
   \frac{1}{L}\log Z_{\rm cyl}(\tilde{\beta_n},\alpha)
   =\frac{\pi c_{m}}{3\tilde{\beta_n}}\,
     \frac{1}{1+\sqrt{1+\frac{\pi c_{m}\alpha}{6\tilde{\beta_n}^{2}}}}\,.
\end{equation}
Differentiating the cylinder answer gives the operator conjugate to
the dressed deformation,
\begin{equation}
\label{chm:cyl-TT-alpha}
   -\frac{1}{L\tilde{\beta_n}}\,
    \frac{\partial\log Z_{\rm cyl}}{\partial\alpha}
   =
   \frac{2}{\alpha^{2}}\!
   \left[
      \frac{1+\frac{\pi c_{m}\alpha}{12\tilde{\beta_n}^{2}}}
           {\sqrt{1+\frac{\pi c_{m}\alpha}{6\tilde{\beta_n}^{2}}}}
      -1
   \right].
\end{equation}
Using \eqref{chm:alpha-def} and then setting
$\tilde\beta_{n}=2\pi n$ gives the all order deformation operator
\begin{equation}
\label{chm:TT-flow}
   \tilde T\tilde{\bar T}_{\rm cyl}(\Phi,n)
   =\frac{2 e^{4\Phi}}{\lambda^{2}}\!
   \left[\frac{1+Y}{\sqrt{1+2Y}}-1\right],
   \qquad
   Y\equiv
   \frac{\pi c_{m}\lambda e^{-2\Phi}}
        {12\tilde\beta_{n}^{2}}
   =\frac{\ell^{2}}{2n^{2}}\,e^{-2\Phi}\,.
\end{equation}

The homogeneous Liouville equation is still the algebraic eq.\footnote{Note that the action in \eqref{chm:S} is defined only at linear order in the deformation, $\lambda$. For the all order $\tilde{T}\tilde{\bar{T}}(\lambda)$ operator in \eqref{chm:TT-flow}, the contribution to the Liouville equation of motion needs to be worked out carefully.  In \cite{Allameh:2025gsa}, this is done by promoting the deformation parameter to a position-dependent function, $\lambda\rightarrow\lambda(x)$, and then by relating the scale transformation of the effective action $W$ generated by $\lambda$, and the transformation generated by the Weyl mode, $\Phi$. The latter captures the contribution of the dressed $\tilde{T}\tilde{\bar{T}}(\lambda)$ term to the $\Phi$ equation of motion. The conclusion of this analysis is that this contribution at all orders in $\lambda$ is still governed by $-8\pi b^2\lambda\tilde{T}\tilde{\bar{T}}(\lambda)e^{-2\Phi}$.}
\begin{equation}
\label{chm:EOM}
   2\,\tilde\Box\Phi
   +8\pi b^{2}\!\left[\mu e^{2\Phi}
                -\lambda\,\tilde T\tilde{\bar T}(\lambda,\Phi)\,e^{-2\Phi}\right]=0.
\end{equation}
Substituting \eqref{chm:TT-flow},
\begin{equation}
\label{chm:Y-eq}
   \frac{K\ell}{2}
   =\frac{1+Y_{*}}{\sqrt{1+2Y_{*}}},
   \qquad K\ell\equiv 2+\lambda\mu .
\end{equation}
Squaring gives
\begin{equation}
\label{chm:Y-quadratic}
   Y_{*}^{2}
   +\left(2-\frac{K^{2}\ell^{2}}{2}\right)Y_{*}
   +\left(1-\frac{K^{2}\ell^{2}}{4}\right)=0.
\end{equation}
With $\Delta=\sqrt{K^{2}\ell^{2}-4}$ the physical root is
\begin{equation}
\label{chm:Y-star}
   Y_{*}
   =\frac{\Delta(K\ell+\Delta)}{4}
   =\frac{\Delta}{K\ell-\Delta}\,,
\end{equation}
while the other root is negative for $K\ell>2$. Therefore
\begin{equation}
\label{chm:Phi-star}
  e^{2\Phi_{*}(n)}
   =\frac{\pi c_{m}\lambda}{12\tilde\beta_{n}^{2}Y_{*}}
   =\frac{\ell^{2}}{2n^{2}}\,
     \frac{K\ell-\Delta}{\Delta}\,.
\end{equation}
Expanding near \(K\ell=2\), or equivalently for small
\(\lambda\mu\), gives
\[
   e^{2\Phi_*}
   =
   \frac{\ell^2}{2n^2}
   \left[
      \frac{1}{\sqrt{\lambda\mu}}
      -1
      +\frac{3}{8}\sqrt{\lambda\mu}
      +O\!\left((\lambda\mu)^{3/2}\right)
   \right].
\]
Thus the leading term agrees with the first-order saddle
\eqref{chm:saddle-1st-final}, while the all-order result generates
non-analytic corrections in powers of \(\sqrt{\lambda\mu}\).

\subsection*{On-shell action}
\label{chm:sec-onshell}

With the homogeneous $\Phi=constant$ ansatz
$(\tilde\nabla\Phi)=0$, the partition function
is obtained by combining the flowed matter partition function
\eqref{chm:cyl-logZ-alpha} with the Liouville cosmological term:
\begin{eqnarray}
    Z_{\rm hom}= Z_m^{T\bar{T}} (\alpha)\times  Z_{\rm liouville}.
\end{eqnarray}
First term is the $T\bar{T}$ flowed matter partition function, while the second term is the Liouville contribution at the homogeneous saddle ansatz.
\begin{equation}
\label{chm:logZ-Phi}
   \log Z_{\rm hom}(\Phi_*,n)
   =
   L\,\frac{\pi c_{m}}{3\tilde\beta_{n}}\,
     \frac{1}{1+\sqrt{1+2Y_*}}
   -L\,\tilde\beta_{n}\,\mu e^{2\Phi_*},
   \qquad L=u_{\max}.
\end{equation}
The first term is written from \eqref{chm:cyl-logZ-alpha} while the second contribution comes from the Liouville fields evaluated at the saddle $\Phi_*$.
For sanity check, Extremising \eqref{chm:logZ-Phi} with respect to $\Phi$ reproduces
\eqref{chm:saddle-algebraic}: the derivative of the flowed matter
term is
$2L\tilde\beta_{n}\lambda e^{-2\Phi}
\tilde T\tilde{\bar T}_{\rm cyl}$, while the Liouville term contributes
$-2L\tilde\beta_{n}\mu e^{2\Phi}$.\\

At the saddle, $\sqrt{1+2Y_{*}}=(K\ell+\Delta)/2$ and
\eqref{chm:Phi-star} gives
\begin{align}
\label{chm:onshell-density}
   \frac{1}{L}\log Z_{\rm hom}(n)
   &=
   \frac{\pi c_{m}}{3\tilde\beta_{n}}\,
     \frac{1}{1+\frac{K\ell+\Delta}{2}}
   -\tilde\beta_{n}\mu
     \frac{\ell^{2}}{2n^{2}}\frac{K\ell-\Delta}{\Delta}
     \nonumber\\
   &=
   \frac{\pi c_{m}}{12\tilde\beta_{n}}\,
      (K\ell-\Delta).
\end{align}
Thus, with $\tilde\beta_{n}=2\pi n$,
\begin{equation}
\label{chm:logZ-const-final}
   \log Z_{\rm hom}(n)
   =\frac{c_{m}u_{\max}}{12n}\,
     \frac{K\ell-\Delta}{2}
   =\frac{\ell u_{\max}}{16Gn}(K\ell-\Delta)\,.
 \end{equation}

\subsection*{Replica derivative of the homogeneous saddle}
\label{chm:sec-constant-diagnostic}
 Since \eqref{chm:logZ-const-final} has the form $A/n$,
\begin{equation}
\label{chm:replica-deriv}
   S_{\rm hom}
   =\left.(1-n\partial_{n})\log Z_{\rm hom}(n)\right|_{n=1}
   =2A .
\end{equation}
But we have two end points of the regions at $-\phi_0 $ and $\phi_0$, and each contributes equally, hence we have 
\begin{equation}
\label{chm:SEE-const}
 S_{\rm EE}^{\rm hom}
   =\frac{c_{m}u_{\max}}{3}\,
     \frac{K\ell-\Delta}{2}
   =\frac{\ell u_{\max}}{4G}(K\ell-\Delta)\,.
\end{equation}

Near $K\ell=2$ one has
$(K\ell-\Delta)/2=1-\sqrt{K\ell-2}+O(K\ell-2)$, so the result reduces
to the CFT answer with the leading deformation correction obtained from the cylinder flow. In summary, the entanglement entropy with ($\delta\phi\ll 1$ with UV cutoff $\epsilon= R \delta \phi$) can be written as
\begin{equation}
\label{eq:finaleeans}
\boxed{
    S_{\rm EE}= \frac{c_{\rm eff}}{3} 
\ln\!\left(\frac{2\sin\phi_0}{\delta\phi}\right)=\frac{c_{\rm eff}}{3} 
\ln\!\left(\frac{2 R \sin\phi_0}{\epsilon}\right).}
\end{equation}
Here, we see that the entanglement entropy for a region on a cylinder in a vacuum state is governed by $c_{\rm eff}$. As emphasized earlier, we shouldn't compare this to the bulk result, since these are two different sets of calculations: one in the bulk (global AdS) and the other at the boundary CFT in the vacuum state. 


In any case, we can take the $K \ell \rightarrow 2$ limit of the bulk answer, and in that limit, the arcsinh of $u_{\rm max}$ becomes the logarithm \eqref{eq:finaleeans} that we have above. In this limit, the Liouville potential vanishes $\mu \sim K \ell -2 \rightarrow 0$. Then, the saddle point for the Liouville field $ e^{4\Phi_{*}^{(1)}}=\frac{\lambda\mathcal T}{\mu}$ gives $\Phi_* \rightarrow \infty$. Hence, the Liouville fields decouple completely, leaving us with matter-field entropy.


\section{Discussion}
\label{discussion}
Entanglement entropy (EE) is an important observable in AdS/CFT. In most cases, it is much easier to compute the EE using holography, i.e., the area of the minimal surface divided by $4 G_N$. In this work, we have studied the entanglement entropy of theories obeying conformal boundary conditions. The dual field theory \cite{Allameh:2025gsa} 
consists of an ordinary holographic CFT, which we refer to as the matter theory, coupled to Liouville theory and deformed by a marginal operator of the form $T\bar T e^{-2\xi\Phi}$.  The data of the conformal boundary, in particular the trace of the extrinsic curvature, enters through the Liouville potential, schematically $\mu=(K\ell-2)/16 \pi G \ell $.

Our results show that the RT formula is not modified by conformal boundary conditions: the entanglement entropy of a subregion is still the area of the minimal bulk surface anchored on the entangling surface divided by $4 G_N$. In particular, the unfixed Weyl mode at the boundary does not affect the result. For global AdS$_3$ we find:

\begin{equation}\label{eq:diss_Sa_cbc}
S_A = \frac{c_m}{3}\,\mathrm{arcsinh}\!\left[\sqrt{\frac{K\ell-\Delta}{2\Delta}}\,\sin\phi_0\right], \qquad \Delta = \sqrt{K^2\ell^2-4}.\end{equation}

We  also
performed the CFT analysis on the cylinder, and calculated the EE in vacuum state for the dual field theory. From this calculation we find,
\begin{eqnarray}\label{eq:diss_SEE_conf_bdy}
     S_{\rm EE}=\frac{c_{\rm eff}}{3} 
\ln\!\left(\frac{2 R \sin\phi_0}{\epsilon}\right).
\end{eqnarray}
Note that in this case, the entropy is governed by the effective central charge $c_{\rm eff}$ not by $c_m.$ This is in agreement with 
 \cite{Allameh:2025gsa} where the high-temperature density of states was also found to be governed by $c_{\rm eff}$. The above formula shows the familiar divergence, consistent with the UV locality. As the dual theory is non-unitary and contains states with $h_{\ rmmin}<0$, the vacuum state on the dual theory wouldn't correspond to global AdS (minimal Brown-York energy).
 
There is an interesting limit of our BTZ analysis \eqref{masslessbtz} we would like to point out. We took the $m=0$ limit of the BTZ black hole. In that limit, the metric becomes the Poincaré patch of AdS, which has zero Brown-York energy. Even though this case is somewhat degenerate, since every finite $ r$-hypersurface has $K \ell=2,$ there are  some interesting features to emphasize. The entanglement entropy of a subregion is given by $S_A=\frac{c_m}{3}\log \left( \frac{L}{\epsilon} \right)$, where $L$ is the length of the interval. We can apply a conformal transformation to map a state on the plane to the cylinder. The bulk geometry dual to the state on the cylinder will have zero energy as the anomaly central charge vanishes. We can compare the resulting entropy on the cylinder to \eqref{eq:diss_SEE_conf_bdy} in the $K \ell=2$ limit as $c_{\rm eff} \rightarrow c_m$.



The finiteness of the entanglement at fixed $K$ deserves a comment. In a local continuum QFT, the entanglement entropy of a spatial region is generically UV divergent, so a finite answer might at first look like evidence against a local interpretation of the dual. However, one reading, which we find reasonable, is that the theory on the cutoff surface is an effective theory with a finite UV cutoff set by $ (K\ell -2) $. From this point of view, there is no tension at all; any effective theory at a fixed cutoff has finite entanglement entropy and correlation functions.

 A complementary perspective makes no appeal to that question at all. Rather than treating the fixed-$K$ theory as a cutoff version of some asymptotic
CFT awaiting a UV completion, one may ask about a duality directly between a
finite region of AdS, bounded by a surface of fixed $K$, and a theory defined intrinsically on that surface, with no reference to the asymptotic
boundary anywhere in the statement. In this framing the boundary theory is
not an approximation to anything but  a self-contained system in a finite
box. And the finiteness of its entanglement entropy is then simply what one
expects of such a system, in straightforward agreement with the bulk results. 

Let us also comment on $c_{\rm eff}$. We found that the $c_{\rm eff}$ of
\cite{Allameh:2025gsa} also controls the logarithmic growth of the
entanglement entropy with subregion size. One might worry that since the
undeformed matter theory is a CFT with central charge $c_m$, and an exactly
marginal deformation should not change the central charge, the entropy ought
to be governed by $c_m$ rather than $c_{\rm eff}$. The point is that the
protected quantity is the anomaly coefficient, which is
unchanged (indeed it vanishes  \cite{Allameh:2025gsa}), whereas
$c_{\rm eff}$ is the effective central charge governing the density of states.
In a non-unitary theory, the two differ, $c_{\rm eff}=c-24\,h_{\min}$, and our
boundary theory is non-unitary, since the coupling to timelike Liouville theory
introduces operators of negative dimension below the identity. 
One caveat worth pointing out is that this exact marginality is established only semiclassically \cite{Allameh:2025gsa}. A sharper handle on the
interpretation of $c_{\rm eff}$, and on the nature of the dual theory, would
come from studying correlation functions of the conformal Brown-York stress
tensor in the spirit of \cite{Kraus:2018xrn} and asking whether $c_{\rm eff}$ can be recovered
from them. On the sphere, these $n$-point functions all vanish
\cite{Allameh:2025gsa}, consistent with the vanishing anomaly charge but
insensitive to an effective central charge. It would be interesting
to repeat the analysis on a cylinder or torus and see if  $c_{\rm eff}$ appears.

We close by collecting other open questions and future directions we find exciting.\\

\textbf{Studying the dual CFT.} One important direction is to better understand the dual theory, including its spectrum, OPE coefficients, and operator algebra. This dual theory is dual to the gravity theory with conformal boundary condition. It will be interesting to find out whether OPE coefficients get a potential $K \ell- \sqrt{K^2\ell^2-4}$ correction. There are also subtleties regarding the mapping states from one conformal frame to another. As the total anomaly vanishes, the stress tensor in one conformal frame must transform homogeneously under conformal transformations. Traditionally, the nonhomogeneous term given by the Schwarzian derivative encodes relevant information about the ground state. In bulk, as explained in \cite{Allameh:2025gsa}, global AdS which is dual to the state on a circle, and to get a bulk geometry dual to the state on the plane, one needs to uncompactify the angular direction. But this geometry is singular at $r=0$. 
\begin{eqnarray}
    ds^2= -(1+\frac{r^2}{\ell^2}) dt^2+\frac{dr^2}{(1+\frac{r^2}{\ell^2})}+r^2 dx^2.
\end{eqnarray}
To obtain the Poincaré patch, one need to take the massless limit of the BTZ black hole. The large temperature density of states also gets a correction proportional to Liouville coupling, or in bulk language, $K \ell- \sqrt{K^2\ell^2-4}$. It would be interesting to understand the operators and their conformal dimension in this setup.

It would also be very interesting to compute the entanglement entropy using twist operators, following the approach of Cardy et al. \cite{Cardy:1986ie,Calabrese:2004eu,Calabrese:2005in,Calabrese:2006rx,Calabrese:2009ez,Calabrese:2009qy}. In the present setting, the twist operators are expected to be dressed by Liouville fields, and understanding their correlation functions in the deformed theory as done in the original Liouville theory by \cite{Dorn:1994xn,Zamolodchikov:1995aa} could provide an independent and complementary perspective on the CFT analysis carried out in this article. Furthermore, it would be interesting to establish prescriptions on both sides of the duality for other entanglement measures, such as reflected entropy and pseudo-entropy. For $T\bar{T}$ deformed CFTs, these quantities have been studied in some interesting scenarios \cite{Basu:2024bal,He:2023wko}.\\

\textbf{Formalizing the AdS/CFT dictionary for CBC.} A broader question is whether the familiar holographic dictionary can be generalized to incorporate conformal boundary conditions. The standard AdS/CFT dictionary is built on the asymptotic expansion of bulk fields near the conformal boundary, i.e. in the limit where the radial coordinate approaches the boundary ($z \to 0$ in 
Poincaré coordinates). In this limit, a scalar field dual to an operator of dimension $\Delta$ admits two independent asymptotic behaviors, $\phi \sim \alpha\, z^{\,d-\Delta} + \beta\, z^{\,\Delta}$. The non-normalizable (leading) mode $z^{\,d-\Delta}$ fixes the source for the dual operator, while the normalizable (subleading) mode $z^{\,\Delta}$ encodes its expectation value, the one-point function $\langle \mathcal{O} \rangle$. At a finite radial cutoff, however, the solution no longer separates cleanly into these two power-law modes. So identifying the appropriate source and expectation value on the cutoff surface becomes considerably more subtle. Conformal boundary conditions fix the conformal class of the boundary metric together with the trace of the extrinsic curvature. Since this is a genuinely different boundary value problem from the standard Dirichlet one, the boundary generating functional, and hence the correlation functions of the dual operators, are modified relative to the undeformed CFT. 
From the field-theory side, correlation functions are known to flow 
under the $T\bar{T}$ deformation. It would therefore be interesting to compute these deformed correlators on both sides of the duality, directly in the boundary theory and holographically from the bulk.\\

\textbf{Studying Black-hole interior:-}
An interesting aspect of the conformal boundary condition is that the cutoff surface can be located anywhere in the bulk of AdS. In this article, we have studied the cutoff surface outside the horizon, $r_c > r_h$, which corresponds to $K \ell > 2$. But one can consider having the cutoff surface inside the horizon or at the stretched horizon. It will change the sign of $K$, hence flipping the sign of the Liouville coupling. This leads to unbounded potential and may be relevant for understanding the bulk geometry in the interior. It would be interesting to understand all these cases, including extremal black holes in 3d and their higher-dimensional counterparts. See \cite{Taylor:2018xcy, Banerjee:2019ewu} for related work in $T\bar{T}$ and entanglement entropy in higher dimensions.

There has been some work in more general boundary conditions for AdS/CFT in which boundary conditions are treated as part of holographic data and may evolve under radial RG flow \cite{Parvizi:2025shq, Parvizi:2025wsg}. It would be interesting to understand the interior of a black hole from this perspective. It would also be interesting to connect the conformal boundary condition perspective to the Cauchy slice holography of Wall et.al. \cite{Araujo-Regado:2022gvw,Soni:2024aop}.\\


\textbf{Understanding other Holographic observables:-} As we have seen the conformal boundary condition changes the Gibbons-Hawking-York term. With this change, it would be interesting to examine other observables, such as the complexity-action or complexity-volume conjectures. One would also like to understand the contribution of the joints where two surfaces meet. One can set up a variational principle that respects the conformal boundary condition in such cases as well. \\


\textbf{CBC for null boundaries:- }Much of the work on conformal boundary conditions has been done for the case of timelike or spacelike boundaries. CBC for Einstein gravity with vanishing cosmological constant has been studied in \cite{Banihashemi:2024yye}. Since the holographic dual of asymptotically flat spacetimes has been conjectured to live on the future/past null infinity, it would be interesting to see how conformal boundary conditions can be applied to a null boundary. One can explore whether the boundary term in the action needs to be suitably modified and whether the conjectured dual Carrollian CFT holds.

We hope to return to some of these questions in future work.

\section*{Acknowledgements}
We want to thank Sanjit Shashi for his collaboration in the initial stages of this project and for many useful conversations. We would also like to thank D. Galante, A. Karch,  E. Shaghoulian for enlightening discussions and comments on the earlier versions of the draft. The work of E.C. and H.K.  is supported in part by CNS Spark Grant 2025.
\appendix

\section{Non-rotating BTZ entropy details}
\label{calculation}

In this appendix, we present explicit details of the calculation of subregion entropy for a non-rotating BTZ black hole background.  We also present the result in terms of the $c_{\text{eff}}$ found by \cite{Allameh:2025gsa} and show that the leading order term matches with the expected Cardy term.

According to \eqref{subregion entropy BTZ}, the geodesic length calculation (assuming RT) gives the following subregion entanglement entropy (we set $\ell=1$):
\begin{equation}
S=\frac{1}{2G_N}\log\!\left[
\sqrt{1+\left(\frac{\beta r_c}{2\pi}\right)^2
\sinh^2\!\left(\frac{2\pi h}{\beta}\right)}
+\frac{\beta r_c}{2\pi}\sinh\!\left(\frac{2\pi h}{\beta}\right)
\right]
\end{equation}
where, $\beta$ is the temperature as observed at the asymptotic boundary, and the subregion is at a finite boundary at $r_c$ between the transverse coordinate points $-h$ to $h$.

It is convenient to define
\[
x\equiv \frac{\beta r_c}{2\pi}=\frac{r_c}{\sqrt m},
\qquad
L_A\equiv 2h,
\qquad
\Delta\equiv \sqrt{K^2\ell^2-4}.
\]
Then
\[
S=\frac{1}{2G_N}\log\!\left[
\sqrt{1+x^2\sinh^2\!\left(\frac{\pi L_A}{\beta}\right)}
+x\,\sinh\!\left(\frac{\pi L_A}{\beta}\right)
\right].
\]

Calculating the trace of extrinsic curvature of the $r=r_c$ hypersurface gives:
\[
K=\frac{2r_c^2-m}{r_c\sqrt{r_c^2-m}}
\]
and we rewrite \(K\) in terms of \(x\) as
\[
K=\frac{2x^2-1}{x\sqrt{x^2-1}}
\]
Squaring and solving for \(x^2\) gives
\[
(K^2-4)x^4+(4-K^2)x^2-1=0,
\]
so that
\[
x^2=\frac12\left(1\pm \frac{K}{\sqrt{K^2-4}}\right)
=\frac12\left(1\pm \frac{K}{\Delta}\right).
\]
Since \(x=r_c/\sqrt m>1\), we must choose the branch
\[
x^2=\frac12\left(1+\frac{K}{\Delta}\right)
=\frac{K+\Delta}{2\Delta},
\qquad
x=\sqrt{\frac{K+\Delta}{2\Delta}}.
\]

Therefore
\[
S(K,L_A,\beta)
=
\frac{\ell}{2G_N}\log\!\left[
\sqrt{1+\frac{K\ell+\Delta}{2\Delta}\,
\sinh^2\!\left(\frac{\pi L_A\ell}{\beta}\right)}
+\sqrt{\frac{K\ell+\Delta}{2\Delta}}\,
\sinh\!\left(\frac{\pi L_A\ell}{\beta}\right)
\right],
\]

where we have re-introduced $\ell$.
Equivalently, using
\[
\log\!\bigl(\sqrt{1+u^2}+u\bigr)=\operatorname{arcsinh}(u),
\]
this is
\[
S(K,L_A,\beta)
=
\frac{\ell}{2G_N}\,
\operatorname{arcsinh}\!\left[
\sqrt{\frac{K\ell+\Delta}{2\Delta}}\,
\sinh\!\left(\frac{\pi L_A\ell}{\beta}\right)
\right].
\]

Now we use the conformal-temperature ($\tilde\beta$) for non-rotating BTZ, which is given by:
\[
\tilde\beta
=
\beta\,\frac{\sqrt{r_c^2-m}}{\ell r_c}
=
\beta\,\frac{\sqrt{x^2-1}}{\ell x}.
\]
Since
\[
x^2=\frac{K\ell+\Delta}{2\Delta},
\qquad
x^2-1=\frac{K\ell-\Delta}{2\Delta},
\]
we get
\[
\frac{\tilde\beta}{\beta}
=
\frac{1}{\ell}\sqrt{\frac{x^2-1}{x^2}}
=
\sqrt{\frac{K\ell-\Delta}{K\ell+\Delta}}
=
\frac{K\ell-\Delta}{2},
\]
where in the last step we used
\[
(K\ell-\Delta)(K\ell+\Delta)=4.
\]
Hence
\[
\beta=\frac{2\tilde\beta\ell}{K\ell-\Delta}.
\]

Substituting this into \(S(K,L_A,\beta)\), we obtain
\[
S(K,L_A,\tilde\beta)
=
\frac{\ell}{2G_N}\,
\operatorname{arcsinh}\!\left[
\sqrt{\frac{K\ell+\Delta}{2\Delta}}\,
\sinh\!\left(
\frac{\pi (K\ell-\Delta)}{2\tilde\beta}\,L_A
\right)
\right].
\]


Now let
\[
Y\equiv \frac{\pi (K\ell-\Delta)}{2\tilde\beta}\,L_A.
\]
We consider the case when the subregion size is large compared to the inverse temperature i.e., $L_A\gg\beta$. Then (\(Y\gg 1\)) which implies
\[
\sinh Y\sim \frac12 e^Y
\]
Since we are looking at conformal boundaries outside the black hole horizon ($r_c>r_H$) the factor $\sqrt{(K\ell+\Delta)/2\Delta}=r_c/\sqrt{m}$ is larger than 1. And for any large $z$: 
\[
\operatorname{arcsinh}(z)\sim \log(2z)
\]
Therefore
\[
S(K,L_A,\tilde\beta)
\sim
\frac{Y\ell}{2G_N}
+\frac{\ell}{2G_N}\log\!\sqrt{\frac{K\ell+\Delta}{2\Delta}}.
\]
which gives
\[
S(K,L_A,\tilde\beta)
\sim
\frac{\pi\ell (K\ell-\Delta)}{4G_N\,\tilde\beta}\,L_A
+\frac{\ell}{4G_N}\log\!\left(\frac{K\ell+\Delta}{2\Delta}\right),
\qquad L_A\to\infty.
\]
Now we use
\[
c_{\rm eff}
=
\frac{3\ell}{4G_N}(K-\Delta).
\]
Then the leading term becomes
\[
S_{\rm leading}(L_A,\tilde\beta)
=
\frac{\pi c_{\rm eff}}{3\tilde\beta}\,L_A.
\]
For the subleading term, first note that
\[
K\ell-\Delta=\frac{4G_N}{3}c_{\rm eff},
\qquad
K\ell+\Delta=\frac{4}{K\ell-\Delta}=\frac{3}{G_N c_{\rm eff}}.
\]
Hence
\[
\Delta=\frac12\Bigl((K\ell+\Delta)-(K\ell-\Delta)\Bigr)
=
\frac{3}{2G_N c_{\rm eff}}-\frac{2G_N c_{\rm eff}}{3},
\]
and therefore
\[
\frac{K\ell+\Delta}{2\Delta}
=
\frac{1}{1-\frac{4G_N^2 c_{\rm eff}^2}{9}}.
\]
So the subleading term can be written purely in terms of \(c_{\rm eff}\) as
\[
S_{\rm subleading}
=
-\frac{\ell}{4G_N}
\log\!\left(1-\frac{4G_N^2 c_{\rm eff}^2}{9}\right).
\]

Thus the final large-\(\tilde{\beta}\) expansion is
\[
S(L_A,\tilde\beta)
\sim
\frac{\pi c_{\rm eff}}{3\tilde\beta}\,L_A
-\frac{\ell}{4G_N}
\log\!\left(1-\frac{4G_N^2 c_{\rm eff}^2}{9}\right)
\qquad (\tilde{\beta}\to\infty).
\]


\section{Rotating BTZ entropy calculation}
\label{rotating_cal}

In this appendix we calculate the subregion entanglement entropy in the background of a rotating BTZ black hole (assuming RT formula) following a similar line of calculations as the Appendix \ref{calculation}. Again we are assuming conformal boundary conditions so the boundary is at a finite $r=r_c$.



We consider the two endpoints of the interval \footnote{Here, the length of the interval is $l$ to simplify the expressions in subsequent equations. Length of the interval is $l=L_A$.} at
\[
t=0,\qquad x=\pm \frac{l}{2},\qquad r=r_c.
\]

The rotating BTZ \(\to\) Poincaré map is
\[
w_{\pm}
=
\sqrt{\frac{r^2-r_+^2}{r^2-r_-^2}}\,e^{(x\pm t)(r_+\pm r_-)},
\qquad
z=
\sqrt{\frac{r_+^2-r_-^2}{r^2-r_-^2}}\,e^{xr_+ + tr_-}.
\]

Define
\[
A\equiv \sqrt{\frac{r_c^2-r_+^2}{r_c^2-r_-^2}},
\qquad
B\equiv \sqrt{\frac{r_+^2-r_-^2}{r_c^2-r_-^2}}.
\]
Then the two endpoints map to

\[
P_1:\qquad
w_{+,1}=A\,e^{-\frac{l}{2}(r_+ + r_-)},
\qquad
w_{-,1}=A\,e^{-\frac{l}{2}(r_+ - r_-)},
\qquad
z_1=B\,e^{-\frac{r_+l}{2}},
\]

\[
P_2:\qquad
w_{+,2}=A\,e^{+\frac{l}{2}(r_+ + r_-)},
\qquad
w_{-,2}=A\,e^{+\frac{l}{2}(r_+ - r_-)},
\qquad
z_2=B\,e^{+\frac{r_+l}{2}}.
\]

Now we use the Poincaré AdS geodesic-distance invariant $L$ for two spacelike-separated points:
\[
\cosh L
=
\frac{(w_{+,1}-w_{+,2})(w_{-,1}-w_{-,2})+z_1^2+z_2^2}{2z_1z_2}.
\]

We first compute the differences:
\[
w_{+,1}-w_{+,2}
=
-2A\,\sinh\!\left(\frac{(r_+ + r_-)l}{2}\right),
\]
\[
w_{-,1}-w_{-,2}
=
-2A\,\sinh\!\left(\frac{(r_+ - r_-)l}{2}\right).
\]
Therefore
\[
(w_{+,1}-w_{+,2})(w_{-,1}-w_{-,2})
=
4A^2
\sinh\!\left(\frac{(r_+ + r_-)l}{2}\right)
\sinh\!\left(\frac{(r_+ - r_-)l}{2}\right).
\]

Using the identity
\[
2\sinh u\,\sinh v=\cosh(u+v)-\cosh(u-v),
\]
with
\[
u=\frac{(r_+ + r_-)l}{2},
\qquad
v=\frac{(r_+ - r_-)l}{2},
\]
we get
\[
(w_{+,1}-w_{+,2})(w_{-,1}-w_{-,2})
=
2A^2\bigl[\cosh(r_+l)-\cosh(r_-l)\bigr].
\]

Next,
\[
z_1^2+z_2^2
=
B^2\left(e^{-r_+l}+e^{r_+l}\right)
=
2B^2\cosh(r_+l),
\]
and
\[
2z_1z_2=2B^2.
\]

Substituting into the geodesic-distance formula, we obtain
\[
\cosh L
=
\frac{2A^2\bigl[\cosh(r_+l)-\cosh(r_-l)\bigr]+2B^2\cosh(r_+l)}{2B^2}.
\]
Hence
\[
\cosh L
=
\left(1+\frac{A^2}{B^2}\right)\cosh(r_+l)-\frac{A^2}{B^2}\cosh(r_-l).
\]

Now,
\[
\frac{A^2}{B^2}
=
\frac{\frac{r_c^2-r_+^2}{r_c^2-r_-^2}}{\frac{r_+^2-r_-^2}{r_c^2-r_-^2}}
=
\frac{r_c^2-r_+^2}{r_+^2-r_-^2},
\]
so
\[
1+\frac{A^2}{B^2}
=
\frac{r_+^2-r_-^2+r_c^2-r_+^2}{r_+^2-r_-^2}
=
\frac{r_c^2-r_-^2}{r_+^2-r_-^2}.
\]

Therefore the geodesic length satisfies
\[
\cosh L
=
\frac{r_c^2-r_-^2}{r_+^2-r_-^2}\cosh(r_+l)
-
\frac{r_c^2-r_+^2}{r_+^2-r_-^2}\cosh(r_-l).
\]

Finally, the holographic entanglement entropy of the subregion is
\[
S_{\rm rot}=\frac{L}{4G}
=
\frac{1}{4G}
\operatorname{arccosh}\!\left[
\frac{r_c^2-r_-^2}{r_+^2-r_-^2}\cosh(r_+l)
-
\frac{r_c^2-r_+^2}{r_+^2-r_-^2}\cosh(r_-l)
\right].
\]

Now we want to write this in terms of $\tilde\beta,\ K,\ \tilde\Omega,\ l$ and see the large $l$ limit ($l\gg \beta$).
We define
\[
\Delta \equiv \sqrt{K^2\ell^2-4},
\qquad
a \equiv K\ell-\Delta.
\]
Then 
\[
r_+=\frac{\pi a\ell}{\tilde\beta(1-\tilde\Omega^2)},
\qquad
r_-=\frac{2\pi\ell\tilde\Omega}{\tilde\beta(1-\tilde\Omega^2)},
\qquad
r_c=\frac{\pi\ell}{\tilde\beta\sqrt{1-\tilde\Omega^2}}\sqrt{\frac{2a}{\Delta}}.
\]
which gives
\[
\frac{r_c^2-r_-^2}{r_+^2-r_-^2}
=
\frac{K\ell+\Delta}{2\Delta},
\qquad
\frac{r_c^2-r_+^2}{r_+^2-r_-^2}
=
\frac{K\ell-\Delta}{2\Delta}.
\]
Hence
\[
S_{\rm rot}(K,\tilde\beta,\tilde\Omega,l)
=
\frac{\ell}{4G}
\operatorname{arccosh}\!\left[
\frac{K\ell+\Delta}{2\Delta}\,
\cosh\!\left(
\frac{\pi (K\ell-\Delta)}{\tilde\beta(1-\tilde\Omega^2)}\,l
\right)
-
\frac{K\ell-\Delta}{2\Delta}\,
\cosh\!\left(
\frac{2\pi \tilde\Omega}{\tilde\beta(1-\tilde\Omega^2)}\,l
\right)
\right].
\]


Since $r_+>r_-$, the \(r_+\)-term dominates at large \(l\). We know
\[
\operatorname{arccosh}(y)\sim \log(2y)
\qquad (y\gg 1)
\]
Therefore
\[
S_{\rm rot}(K,\tilde\beta,\tilde\Omega,l)
\sim
\frac{\ell}{4G}
\log\left[\left(\frac{K\ell+\Delta}{2\Delta}\right)2\cosh\left(
\frac{\pi (K\ell-\Delta)}{\tilde\beta(1-\tilde\Omega^2)}\,l\right)
\!
\right]
\]
which gives
\[
S_{\rm rot}(K,\tilde\beta,\tilde\Omega,l)
\sim
\frac{\pi\ell (K\ell-\Delta)}{4G\,\tilde\beta(1-\tilde\Omega^2)}\,l
+
\frac{\ell}{4G}\log\!\left(\frac{K\ell+\Delta}{2\Delta}\right),
\qquad
l\to\infty.
\]


Now we want to see this in terms of $c_{\text{eff}}$. We know
\[
c_{\rm eff}=\frac{3\ell}{4G}(K-\Delta),
\]
then leading term becomes
\[
S_{\rm leading}
=
\frac{\pi c_{\rm eff}}{3\,\tilde\beta(1-\tilde\Omega^2)}\,l.
\]

Also,
\[
K\ell-\Delta=\frac{4G}{3}c_{\rm eff},
\qquad
(K\ell-\Delta)(K\ell+\Delta)=4,
\]
so
\[
\frac{K\ell+\Delta}{2\Delta}
=
\frac{1}{1-\frac{(K\ell-\Delta)^2}{4}}
=
\frac{1}{1-\frac{4G^2 c_{\rm eff}^2}{9}}.
\]
Hence, the subleading term is
\[
S_{\rm subleading}
=
-\frac{\ell}{4G}
\log\!\left(
1-\frac{4G^2 c_{\rm eff}^2}{9}
\right).
\]

Therefore, the final large-\(\tilde{\beta}\) expansion is
\[S_{\rm rot}(\tilde{\beta},\tilde{\Omega},L_A)
\sim
\frac{\pi c_{\rm eff}}{3\,\tilde\beta(1-\tilde\Omega^2)}\,L_A
-\frac{\ell}{4G}
\log\!\left(
1-\frac{4G^2 c_{\rm eff}^2}{9}
\right),
\qquad
\tilde{\beta}\to\infty.\]
\section{Consistency of the CHM frame}
\label{app:chm frame consistency}

In this appendix, we discuss the transformation of the Liouville sector under the CHM map.
Since the Liouville sector is sensitive to the choice of fiducial frame, one can show that the passage to the CHM frame doesn't change the action
(matter-timelike-Liouville theory). Let \(D\) denote the causal-diamond frame on
the cylinder and \(H\) the hyperbolic thermal frame. After the CHM (conformal) transformation, which trivially preserves the action, the two fiducial metrics and Liouville fields are related by
\begin{equation}
    \widetilde g^{(D)}_{ab}
    =
    e^{2\sigma}\,\widetilde g^{(H)}_{ab},
    \qquad
    \sigma \equiv \log\Omega,
    \qquad
    \Phi_D = \Phi_H - \sigma ,
    \label{eq:frame-map}
\end{equation}
so that the physical metric \(e^{2\Phi}\widetilde g_{ab}\) is unchanged or
\begin{equation}
    I\!\left[e^{2\omega}\widetilde g,\;\Phi-\omega\right]
=
I\!\left[\widetilde g,\;\Phi\right].
\end{equation}
It
remains to check invariance under the Weyl factor \(e^{2\sigma}\).

The Liouville potential is separately invariant,
\begin{equation}
    \sqrt{\widetilde g_D}\;\mu\, e^{2\Phi_D}
    =
    \sqrt{\widetilde g_H}\;\mu\, e^{2\Phi_H}.
\end{equation}
For the deformation term, the
\(\widetilde T\widetilde{\overline T}\) denotes the composite built from the stress tensor of the full theory (
matter stress tensor + Liouville kinetic contribution and
excluding the cosmological term). Because the total central charge of this combined system vanishes, this stress tensor transforms homogeneously under Weyl rescalings, with no Schwarzian-type inhomogeneous terms, and the renormalized composite obeys
\begin{equation}
    \bigl(\widetilde T\widetilde{\overline T}\bigr)_D
    =
    e^{-4\sigma}\,
    \bigl(\widetilde T\widetilde{\overline T}\bigr)_H .
    \label{eq:TTbar-weyl}
\end{equation}
The dressed deformation density is frame independent,
\begin{equation}
    \sqrt{\widetilde g_D}\;
    \lambda\,
    \bigl(\widetilde T\widetilde{\overline T}\bigr)_D\,
    e^{-2\Phi_D}
    =
    \sqrt{\widetilde g_H}\;
    \lambda\,
    \bigl(\widetilde T\widetilde{\overline T}\bigr)_H\,
    e^{-2\Phi_H}.
\end{equation}

The kinetic and background-charge terms of Liouville theory
\begin{equation}
    I_{\mathrm{kin+bg}}[\widetilde g,\Phi]
    \equiv
    \frac{1}{4\pi b^2}
    \int d^2x\,\sqrt{\widetilde g}\,
    \Bigl[
        -(\widetilde\nabla\Phi)^2
        -\widetilde R\,\Phi
    \Bigr],
\end{equation}
are not separately Weyl invariant. Under \eqref{eq:frame-map} one finds
\begin{equation}
\begin{split}
    I_{\mathrm{kin+bg}}[\widetilde g_D,\Phi_D]
    &=
    I_{\mathrm{kin+bg}}[\widetilde g_H,\Phi_H]
    \\
    &\quad
    +\frac{1}{4\pi b^2}
    \int d^2x\,\sqrt{\widetilde g_H}\,
    \Bigl[
        (\widetilde\nabla\sigma)^2
        +\widetilde R_H\,\sigma
    \Bigr],
    \label{eq:WZ-shift}
\end{split}
\end{equation}
Since \(b^2 = 6/c_m\), the coefficient of
this Wess--Zumino functional is \(c_m/24\pi\), and it is cancelled by the
anomalous transformation of the matter effective action
\(W_m \equiv -\log Z_m\),
\begin{equation}
    W_m[e^{2\sigma}\widetilde g_H]
    =
    W_m[\widetilde g_H]
    -\frac{c_m}{24\pi}
    \int d^2x\,\sqrt{\widetilde g_H}\,
    \Bigl[
        (\widetilde\nabla\sigma)^2
        +\widetilde R_H\,\sigma
    \Bigr].
\end{equation}
This cancellation uses the classical relation \(b^2=6/c_m\) and therefore
holds at the semiclassical order.

\bibliographystyle{jhep}
\bibliography{refs}

\end{document}